\documentclass[twocolumn]{aastex631}
\usepackage{subfigure}
\usepackage{amsmath}
\usepackage{esint}
\usepackage{color, colortbl}

\definecolor{Gray}{gray}{0.9}
\shorttitle{NGC4486B's Recoiling supermassive Black Hole}
\shortauthors{B. Tahmasebzadeh et al.}
\graphicspath{{./}{figures/}}
\newcommand{\msun}{\mbox{$M_{\odot}$}}
\newcommand{\kms}{\ensuremath{\,\mathrm{km\ s}^{-1}}}

\newcommand{\mbh}{{\cal M}_{BH}}

\newcommand{\vlos}{v_\mathrm{los}}

\begin{document}
%\linenumbers

%\title{A Recent Binary SMBH Merger and Recoil as a Possible Origin of the Double Nucleus in NGC 4486B from JWST Observations}

\title{JWST Observations of the Double Nucleus in NGC 4486B: \\ Possible Evidence for a Recent Binary SMBH Merger and Recoil}

\author[0000-0002-1584-2281]{Behzad Tahmasebzadeh}
\affiliation{Department of Astronomy, University of Michigan,
1085 S. University Ave., Ann Arbor, MI 48109, USA}
\affiliation{Department of Astrophysics and Planetary Science,
Villanova University, 800 East Lancaster Ave., Villanova, PA 18085, USA}

\author[0000-0002-6257-2341]{Monica Valluri}
\affiliation{Department of Astronomy, 
University of Michigan, 1085 S. University Ave., Ann Arbor, MI 48109, USA}

\author[0000-0002-7941-1149]{Shashank Dattathri}
\affiliation{Department of Astronomy, Yale University, Kline Tower 266 Whitney Avenue, New Haven, CT 06511, USA}

\author[0000-0002-0647-718X]{Tatsuya Akiba}
\affiliation{JILA and Department of Astrophysical and Planetary Sciences, CU Boulder, Boulder, CO 80309, USA}

\author[0000-0002-5707-4268]{Fazeel Mahmood Khan}
\affiliation{New York University Abu Dhabi, PO Box 129188, Abu Dhabi, United Arab Emirates}
\affiliation{Center for Astrophysics and Space Science (CASS), New York University Abu Dhabi, PO Box 129188, Abu Dhabi, UAE}

\author[0000-0003-3009-4928]{Matthew A.\ Taylor}
\affiliation{University of Calgary, 2500 University Drive NW, Calgary Alberta T2N 1N4, Canada}

\author[0009-0006-9760-9315]{Haruka Yoshino}
\affiliation{Department of Physics and Astronomy, University of Victoria, PO Box 3055, STN CSC, Victoria BC V8W 3P6, Canada}
\affiliation{National Research Council of Canada, 5071 West Saanich Road, Victoria, BC, V9E 2E7, Canada}

%\author[0000-0002-5038-9267]{Eugene Vasiliev}
%\affiliation{Institute of Astronomy, Madingley Road, Cambridge, CB3 0HA, UK}

\author[0009-0006-7485-7463]{Solveig Thompson}
\affiliation{University of Calgary, 2500 University Drive NW, Calgary Alberta T2N 1N4, Canada}

\author[0000-0002-1119-5769]{Ann-Marie Madigan}
\affiliation{JILA and Department of Astrophysical and Planetary Sciences, CU Boulder, Boulder, CO 80309, USA}

%\author[0000-0003-3234-7247]{Tod R. Lauer}
%\affiliation{NSF National Optical Infrared Astronomy Research Laboratory, P.O. Box 26732, Tucson, AZ 85726, USA}

\author[0000-0003-3236-2068]{Frank C. van den Bosch}
\affiliation{Department of Astronomy, Yale University, Kline Tower 266 Whitney Avenue, New Haven, CT 06511, USA}

\author[0000-0003-2227-1322]{Kelly holley-bockelmann}
\affiliation{Department of Physics and Astronomy, Vanderbilt University, Nashville, TN 37240, USA}
\affiliation{Department of Physics, Fisk University, Nashville, TN 37208, USA}

\author[0000-0003-1184-8114]{Patrick C\^ot\'e}
\affiliation{National Research Council of Canada, 5071 West Saanich Road, Victoria, BC, V9E 2E7, Canada}

\author[0000-0002-8224-1128]{Laura Ferrarese} 
\affiliation{National Research Council of Canada, 5071 West Saanich Road, Victoria, BC, V9E 2E7, Canada}

\author[0000-0003-4867-0022]{Michael J.\ Drinkwater}
\affiliation{School of Mathematics and Physics, University of Queensland,  Brisbane, QLD 4072, Australia}

\author[0000-0002-1959-6946]{Holger Baumgardt}
\affiliation{School of Mathematics and Physics, 
The University of Queensland, St. Lucia, QLD 4072, Australia}

\author[0000-0002-2816-5398]{Misty C.\ Bentz}
\affiliation{Department of Physics and Astronomy, Georgia State University, Atlanta, GA 30303, USA}

\author[0000-0002-8532-4025]{Kristen Dage}
\affiliation{International Centre for Radio Astronomy Research -- Curtin University, GPO Box U1987, Perth, WA 6845, Australia}

\author[0000-0002-2073-2781]{Eric W. Peng}
\affiliation{NSF's National Optical-Infrared Astronomy Research Laboratory, 
950 North Cherry Avenue, 
Tucson, AZ 85719, USA}

\author[0009-0008-1720-3310]{Somya Jha}
\affiliation{Department of Astronomy, 
University of Michigan, 1085 S. University Ave., Ann Arbor, MI 48109, USA}

\author[0000-0002-8171-6507]{Andrea V. Macci\`o}
\affiliation{New York University Abu Dhabi, PO Box 129188, Abu Dhabi, United Arab Emirates}
\affiliation{Center for Astrophysics and Space Science (CASS), New York University Abu Dhabi, PO Box 129188, Abu Dhabi, UAE}

\author[0000-0002-4718-3428]{Chengze Liu}
\affiliation{State Key Laboratory of Dark Matter Physics, Shanghai Key Laboratory for Particle Physics and Cosmology, School of Physics and Astronomy \& Tsung-Dao Lee Institute, Shanghai Jiao Tong University, Shanghai 200240, China}

\author[0000-0003-1428-5775]{Tyrone E.\ Woods}
\affiliation{Department of Physics and Astronomy,
University of Manitoba,
30A Sifton Road,
Winnipeg, Manitoba, R3T 2N2 Canada}

%\author[0000-0002-0363-4266]{Joel Roediger}
%\affiliation{National Research Council of Canada, 
%Herzberg Astronomy and Astrophysics Program, 
%5071 West Saanich Road, 
%Victoria, BC, V9E 2E7, Canada}

%\author[0000-0002-3382-9021]{Kaixiang Wang}
%\affiliation{Department of Astronomy, 
%Peking University, 
%Beijing 100871, People's Republic of China}
%\affiliation{Kavli Institute for Astronomy and Astrophysics, 
%Peking University, 
%Beijing 100871, People's Republic of China}

%\author[0000-0003-4703-7276]{Vivienne Baldassare}
%\affiliation{Department of Physics and Astronomy, Washington State University, Pullman, WA 99163, USA}

%\author[0000-0002-5213-3548]{John P.\ Blakeslee}
%\affiliation{NSF's National Optical-Infrared Astronomy Research Laboratory, 950 North Cherry Avenue, Tucson, AZ 85719, USA}

%\author[0000-0001-6333-599X]{Youkyung Ko}
%\affiliation{Korea Astronomy and Space Science Institute, 776 Daedeok-daero, Yuseong-Gu, Daejeon 34055, Republic of Korea}

\begin{abstract}

A recent study of the compact elliptical galaxy NGC~4486B using JWST–NIRSpec IFU kinematics confirmed a supermassive black hole (SMBH) of mass $\mbh=3.6\pm0.7\times10^8$ ($\sim 8$\% of the stellar mass). In addition to its double nucleus, the nuclear kinematics show pronounced asymmetries: a velocity-dispersion peak displaced by 6 pc from the galaxy center and a $\sim16\kms$ offset in the mean stellar line-of-sight velocity near the SMBH. We examine the origin of the 12 pc double nucleus and show that the observations favor an SMBH surrounded by an eccentric nuclear disk (END). One proposed formation pathway for ENDs involves a gravitational-wave (GW) recoil of the SMBH following a binary merger. Our orbit-superposition models contain $\sim 50 \%$ retrograde stars at the edge of the nuclear region, in agreement with END-formation simulations. We infer a pre-merger mass ratio $q>0.15$ and a recoil kick of $\sim340\kms$. Our $N$-body simulations show that with such a kick the SMBH returns to the center within $\sim30$ Myr. Its flat central core is also consistent with earlier “binary black hole scouring’’. We test two alternative mechanisms—buoyancy-driven oscillations and a pre-merger SMBH binary—but neither reproduces the observed offsets, favoring the GW-kick scenario. Our $N$-body simulations also show that a prograde SMBH binary in a rotating host can become trapped in a corotation resonance, delaying coalescence. Although NGC~4486B is an old galaxy near the Virgo cluster center, its SMBH appears to have merged only recently, making its nucleus a rare nearby laboratory for post-merger SMBH dynamics.

\end{abstract}


%-----------------------------------------------------
\section{Introduction} \label{sec:intro}
NGC~4486B (VCC1279) is a cE galaxy at the center of the Virgo Cluster, projected $7.3^{\prime}$ ($\sim 35$ kpc) from the massive elliptical M87, with a modest line-of-sight velocity difference of $\sim 200 \, \mathrm{km} \, \mathrm{s^{-1}}$ that indicates a close dynamical association. NGC~4486B is characterized by a small effective radius of $2.33^{\prime\prime}$ ($\sim 0.19$ kpc) and an absolute magnitude of $M_V \sim 17.6$, with a total stellar mass of $M_*\sim 6 \times 10^9$\msun\ \citep{Lauer.1996, Ferrarese.2006, Janz.2016}. Recently, we showed using JWST-NIRSpec IFU kinematical data and archival HST photometry, that NGC~4486B  hosts a supermassive black hole (SMBH)  with a mass of $ \mbh = (3.6\pm0.7) \times 10^{8}\msun$, making this black hole significantly overmassive ($\mbh/M_*\sim 8$\%) relative to expectations from black hole scaling relations \citep[][hereafter Paper 1]{Behzad.2025}. This suggests that  NGC~4486B is an extreme case of a tidally stripped galaxy, likely the compact core of a once significantly more massive system.

It has been known since HST first imaged it that NGC~4486B, like M31, has a double nucleus \citep{Lauer.1996}. In NGC 4486B, the brighter peak (P1) and the fainter peak (P2) are separated by $\sim 12$ pc (0.15$^{\prime\prime}$), whereas in M31, the separation is only $\sim 1.9$ pc. The two peaks in NGC 4486B are located at comparable distances from the galaxy's isophotal center (determined by isophotes at large radii), with P1 and P2 each located at $\sim6$ pc. In contrast, the peaks in M31 are much closer to the galaxy's isophotal center, with P1 at $\sim 1.75$ pc and P2 at only $\sim 0.15$ pc. M31 has a third fainter peak P3 at the galactic center \mdash also the location of the SMBH \citep{Lauer.1993}. 

Several formation pathways have been proposed for lopsided or double nuclear structures, including in-situ formation of eccentric disks through gas inflow and self-gravitating $m=1$ instabilities as well as perturbation-driven excitation of long-lived eccentric modes \citep[e.g.,][]{Jacobs.2001, Hopkins.2010}. These mechanisms generally operate in systems with significant gas inflow or massive, self-gravitating nuclear disks. However, NGC~4486B is a gas-poor compact elliptical with an old stellar population and a central potential dominated by its SMBH, making such scenarios less well matched to its present-day properties.

\citet{Lauer.1996} investigated various explanations for the double nucleus in NGC~4486B. A dust absorption feature or an infalling globular cluster were ruled out due to the nearly identical morphologies in the $V$ and $I$ bands and the absence of significant color gradients across the nucleus. A bound stellar binary nucleus was likewise deemed unlikely, as its decay timescale due to dynamical friction would be short ($\lesssim 10^7$ yr), and there is no evidence for a recent galaxy–galaxy merger. Projection effects from unrelated foreground or background sources were also excluded based on the nearly identical colors and recession velocities of the two components \citep{Kormendy.1997}.

One plausible explanation for the double nucleus in M31 and NGC~4486B is the presence of an eccentric nuclear disk (END) \citep{Lauer.1993,Lauer.1996}, apsidally aligned stellar disk orbiting a central SMBH \citep{Tremaine.1995, Peiris.2003, Bacon.2001, Sambhus.2002, BrownMagorrian.2013, Wernke.2021}. In this model, the nearly Keplerian potential of the SMBH dominates the stellar dynamics, shaping a lopsided configuration where stars follow coherent, eccentric orbits. The apparent double nucleus arises from the asymmetric distribution of stellar density along these orbits: the brighter component corresponds to stars near their apocenters, where they move more slowly, while the fainter component is associated with stars near their pericenters, close to the black hole. %\ta{should we mention that we see a double nucleus only for particular inclinations? Only inclinations close to edge-on?}
%(periapses)

A striking feature of NGC4486B is its flat central density core (see Section~\ref{sec:photometry}). One of the best accepted mechanisms for producing such cores in early-type galaxies is via ``binary-black-hole scouring'' following the merger of two gas-poor galaxies, each with its own SMBH.  The black holes (BHs) quickly sink to the common center of the galaxy via dynamical friction and become bound with separations of 10s of parsecs. In order to shrink the orbit further, the SMBH binary transfers its orbital kinetic energy to surrounding stars, via the $3-$body slingshot effect, preferentially ejecting stars with low angular momentum \citep[e.g.][]{Milos.2003, Baile15}  until the binary separation becomes small enough for gravitational wave (GW) radiation to dominate. High-resolution $N$-body simulations \citep{khan+12a,Rantala2024} have shown that an SMBH binary can eject a stellar mass amounting to $2-5$ times its total mass from the central region, thereby converting a central cusp of stars into a flatter central core \citep{rantala2018,Nasim2021,Khan2025}. %\ta{in accordance, my sims to get updated Fig 3 will use a cored density initially}

%\ta{add more citations here} 
Numerical relativity simulations \citep[e.g.,][]{Campanelli.2007ApJ, Campanelli.2007PRL} have shown that the anisotropic emission of GWs during the merger of two SMBH delivers a kick to the final black hole with a magnitude that depends on a variety of factors:
the mass ratio $q = m_2/M_1$  of the secondary to the primary; whether the BHs have spin; the alignment of their spins to their mutual orbital plane; and whether the black hole spins are aligned or misaligned.  While kicks as large as $\sim 2000-4000\kms$ \citep{Campanelli.2007PRL} are possible, the actual distribution of kick velocities is uncertain and most kicks are probably $\lesssim 1000$\kms \citep{Gualandris.2008}. Even a modest kick to the SMBH causes it to oscillate, transferring orbital energy to nearby stars, further flattening the central density profile \citep{Gualandris.2008,Khonji.2024}.% In compact systems such as NGC 4486B, where the potential confines the SMBH near the center, this effect may still reduce the stellar density and imprint subtle phase-space signatures, complementing the impact of binary-driven core scouring.

Recently, the END model has received renewed interest due to a suite of direct $N$-body simulations of nuclear disks surrounding SMBHs  \citep{Akiba.2021, Akiba.2023, Bright.2024}. These authors demonstrate that a GW recoil kick can naturally transform a circular stellar disk bound to an SMBH into an eccentric, apsidally aligned lopsided one. In this scenario, the stars that remain bound to the SMBH after the kick experience angular-momentum changes that place them on eccentric orbits, whose net alignment is oriented roughly orthogonal to the kick direction. For a sufficiently large kick, stars in the outer parts of the disk can even become retrograde in the rest frame of the SMBH.

In this work, we investigate the photometric and kinematic signatures of the double nucleus in NGC~4486B and build upon our previous dynamical modeling results from Paper 1, as well as previous $N$-body simulations, to examine the origin and dynamical state of its END. We use previous analysis from \citet{Akiba.2021, Akiba.2023, Akiba.2025} and the observed properties of the double nucleus to estimate the kick velocity and likely mass-ratio of the black hole binary. We further investigate the dynamical evolution of the central SMBH using a self-consistent $N$-body realization constructed from the Schwarzschild orbit-superposition model in Paper 1. We simulate both the pre-merger evolution of an SMBH binary and the post-merger response of a recoiling SMBH within the host galaxy potential. These experiments probe the large-scale evolution of the SMBH in the global, axisymmetric potential inferred from the data and are not intended to model the internal structure or formation of the END.  We simulate various recoil kicks to assess the timescale on which the black hole would sink to the center. In addition to GW recoil, we test other alternative mechanisms capable of displacing the SMBH and influencing the observed kinematics, such as buoyancy-driven displacements of the SMBH \citep{Cole2012, Banik2021, Banik2022, Dattathri2025a}, and the pre-merger evolution of a prograde binary black hole merger \citep{holley+15,Rasskazov+16,Mirza+17,Khan2020}. We also quantify the expected amplitude of SMBH Brownian motion \citep{Merritt.2007} to verify that the displacements seen in our simulations and inferred from observations exceed the stochastic wandering expected from finite stellar encounters Brownian motion.

While nearly two decades of theoretical and simulation work have explored the consequences of binary black hole mergers, the resulting GW recoils, and SMBH Brownian motion in galaxies, NGC~4486B appears to be the first system exhibiting multiple observable signatures of a recent SMBH merger—making it a benchmark case for studying the dynamical aftermath of SMBH coalescence. Although NGC~4486B and M31 are among the best-studied examples, double nuclei are not unique. \citet{Debattista.2006} examined a sample of dE galaxies from \citet{Lotz.2004} imaged in the WFPC2 snapshot surveys GO-8600 and GO-6352 (PI: H. Ferguson) and found that VCC~1107, FCC~208, and VCC~128 show signs of double nuclei, with VCC~128 exhibiting two comparably bright components and the closest separation ($\sim 32$ pc). %A visual inspection of the NGVS \mv{add ref} also  reveals four galaxies with apparent double nuclei although spectroscopy is required to confirm them \mv{(list galaxies found by Laura)}.
Tracing the formation and evolution of the eccentric disk in NGC~4486B and other galaxies may provide indirect constraints on the frequency of SMBH mergers -- an essential parameter for interpreting the stochastic GW background measured by Pulsar Timing Arrays (PTAs) \citep{Nanograv.2023, nanograv.2024, EPTA.2023, EPTA.2024}.

The structure of this paper is as follows. In Section \ref{sec:data}, we present the HST and JWST observations of NGC~4486B and briefly summarize the Schwarzschild dynamical modeling from Paper~1. In Section \ref{sec:END_recoil}, we present the observational evidence for an END in NGC 4486B and compare the nuclear structure and kinematics of our Schwarzschild model with simulations of ENDs formed through GW recoil kicks. Using the observed properties of the double nucleus, we derive constraints on the black hole binary mass ratio and the required recoil velocity. In Section \ref{ssec:nbody_kick}, we use direct $N$-body simulations to follow the post-kick trajectory of the SMBH and evaluate the timescale on which it returns to the galaxy center, as well as the kinematic signatures such a kick would imprint. Section \ref{sec:altscen} examines alternative mechanisms that could produce an off-center SMBH or similar kinematic asymmetries. Finally, in Section \ref{sec:conclusion} we summarize our findings and discuss their implications. Throughout this study, we adopt a distance of $16.3$ Mpc to NGC 4486B \citep{Blakeslee.2009}.

\section{Observational Evidence and overview of Schwarzschild models in Paper 1} \label{sec:data}

In this section we briefly summarize the properties of photometric imaging from Hubble Space Telescope (HST)  instruments WFPC2 and ACS/WFC and the spectroscopic data from JWST-NIRSpec IFU that were used to construct the Schwarzschild dynamical models of NGC~4486B presented in Paper 1 \citep{Behzad.2025}. In addition, we briefly summarize the main results from that paper, including details about the best-fit Schwarzschild models that are used in this work.
 
\subsection{Photometric data and image analysis \label{sec:photometry}}
\begin{table*}
	\centering
	\footnotesize
	%\begin{tabular}{p{0.1\linewidth}p{0.1\linewidth}p{0.1\linewidth}p{0.1\linewidth}p{0.1\linewidth}p{0.1\linewidth}p{0.1\linewidth}p{0.1\linewidth}}
	\begin{tabular}{cccccccc}
		\hline
		model    &  $n$ &  $r_{e}$ (arcsec) & $\Sigma_{e}$ $\mathrm{mag/arcsec^{2}}$& $q$   &  PA (degree)  & $\chi^{2}/N_{\mathrm{dof}}$ & AIC \\
		\hline
        1 S\'{e}rsic  & 3.05 & 2.78 & 18.58 & 0.79 & 88.67 & 5.5 & 30032248  \\
        \hline
        2 S\'{e}rsic  & 1.2 & 0.78 & 16.45 & 0.65 & 87.90 & 0.67 & 365042  \\
        $\,$          & 1.80 & 2.93 & 18.72 & 0.86 & 89.29 & $\,$ & $\,$  \\
        \hline
        3 S\'{e}rsic  & 1.26 & 0.40 & 16.73 & 0.74 & 90.55 & 0.42 & 227140  \\
        $\,$          & 1.76 & 1.33 & 18.01 & 0.71 & 86.86 & $\,$ & $\,$  \\
        $\,$          & 1.46 & 3.26 & 19.08 & 0.90 & 91.47 & $\,$ & $\,$  \\		
        \hline
        4 S\'{e}rsic  & 1.44 & 0.39 & 17.27 & 0.74 & 100.63 & 0.36 & 192528  \\
        $\,$          & 1.39 & 0.52 & 17.44 & 0.73 & 84.85 & $\,$ & $\,$  \\
        $\,$          & 2.30 & 1.80 & 19.47 & 0.57 & 85.24 & $\,$ & $\,$  \\
        $\,$          & 1.58 & 3.02 & 19.04 & 0.93 & 96.21 & $\,$ & $\,$  \\
        \hline
	\end{tabular}%}
    \\
	\parbox{2\columnwidth}{\caption{The GALFIT output for the best-fit 2D surface brightness models of NGC 4486B F850LP image using one, two, three, and four S\'ersic functions. Columns from left to right list the number of S\'{e}rsic components, S\'{e}rsic indices ($n$), effective radius ($r_{e}$), surface brightness at the effective radius ($\Sigma_{e}$), flattening ($q$), position angle (PA), reduced chi-square ($\chi^{2}/N_{\mathrm{dof}}$, where $N_{\mathrm{dof}}$ is the number of degrees of freedom), and the Akaike Information Criterion (AIC), defined as $\chi^{2} + 2k$, with $k$ representing the number of model parameters for each model.}
		\label{table:galfit}}
\end{table*}
In Paper 1, we used archival  HST-ACS imaging in the F850LP filter (pixel scale $\sim0.05\arcsec$ $\lesssim 4$ pc and PSF FWHM $\sim0.9\arcsec$ $\lesssim 72$ pc) acquired as part of the ACS Virgo Cluster Survey \citep{Cote.2004}. A model point spread function (PSF) was generated using the TinyTim software package \citep{Krist.2011} and used for all photometric analyses. The full photometric frame reaches out to $\sim$1 kpc in radius, with an effective radius of  $2.33\arcsec$ (182 pc), extending well beyond the field of view of the NIRSpec-IFU observations.  We also examine the WFPC2 F555W image from \citet{Lauer.1996}, which provides slightly higher spatial resolution in the central region and reveals the double nucleus structure more distinctly. 

Previous analyses of the morphology, isophotal parameters, and surface brightness profile of NGC 4486B showed that it exhibits a shallow surface brightness profile in its central region \citep{Lauer.1996, Faber.1997, Ferrarese.2006}. \citet{Ferrarese.2006} modeled its surface brightness using a single S\'{e}rsic component and obtained $n = 2.73$ and $r_{e} = 2.33\arcsec$.  We repeated the 2D surface brightness fitting of the ACS F850LP image using GALFIT \citep{GALFIT2010} models with one, two, three, and four S\'{e}rsic components.   For a single component, the S\'{e}rsic index is \( n = 3.05 \), slightly higher than in previous work.  As the number of S\'{e}rsic components increases to three, the quality of the fit improved. Beyond this point, however, the model did not show substantial improvement (see Table~\ref{table:galfit}). By adding more components, the S\'{e}rsic index of the inner region decreases to a range of  $n = 1.2 - 1.4$. As might be expected due to the double nuclei, the inner region of NGC 4486B cannot be adequately described by single or multiple nested S\'{e}rsic components (even allowing for the components to have different centers). The major-axis position angles of all components across different models are similar and consistent with the position angle profile derived from the ellipse fitting to the isophotes.  The ellipticity and position angle of the isophotes (outside the innermost $\sim 0.15\arcsec$ region occupied by the double nucleus) remain relatively constant with radius, indicating that the galaxy is close to axisymmetric overall. Similar results were obtained with the WFPC2 F555W image.
 \begin{figure}
	\centering	%
	\includegraphics[width=\columnwidth]{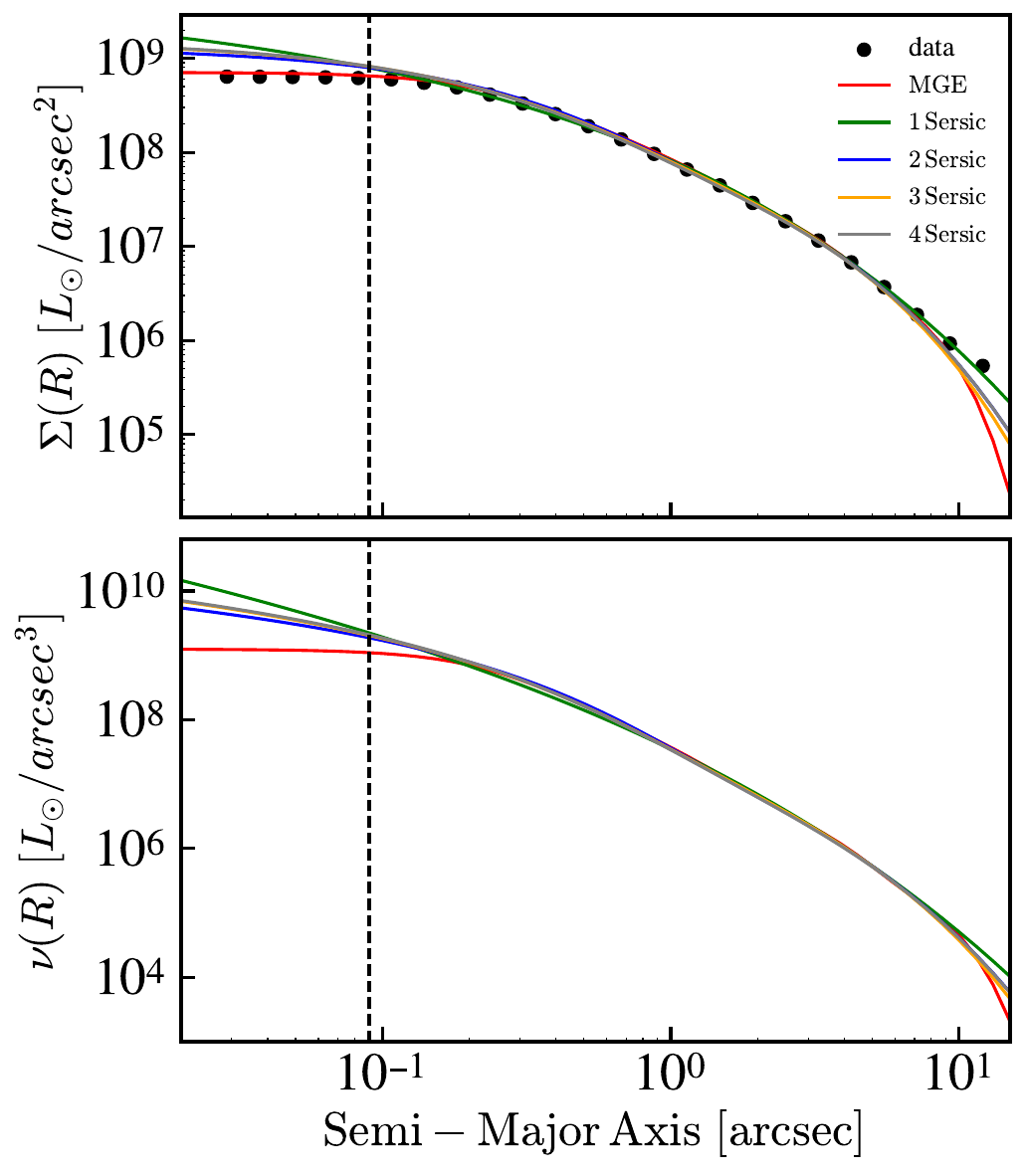}
	\caption{The top panel presents the projected luminosity density of NGC 4486B (black dots) derived from the  ACS F850LP image along with various models (see text for details). The bottom panel displays the deprojected 3D luminosity density profiles along the semi-major axis. The vertical dashed line indicates the PSF FWHM of the ACS F850LP image. }%\mv{check: scale or resolution?}\BT{BT: it is pixel scale}
	\label{fig:3d_profile}%
\end{figure}

 \begin{figure*}[t]
	\centering	%
	\includegraphics[width=2\columnwidth]{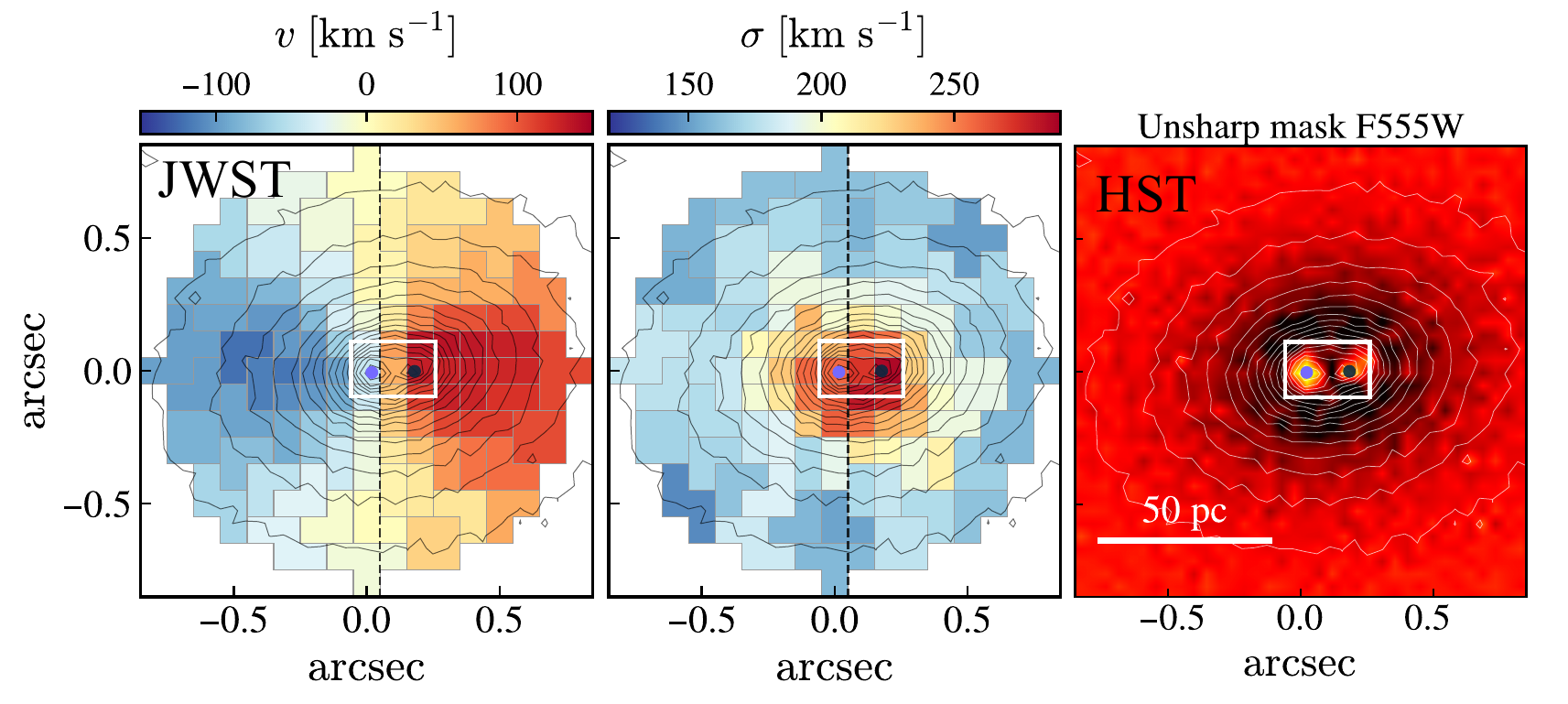}
    \includegraphics[width=2\columnwidth]{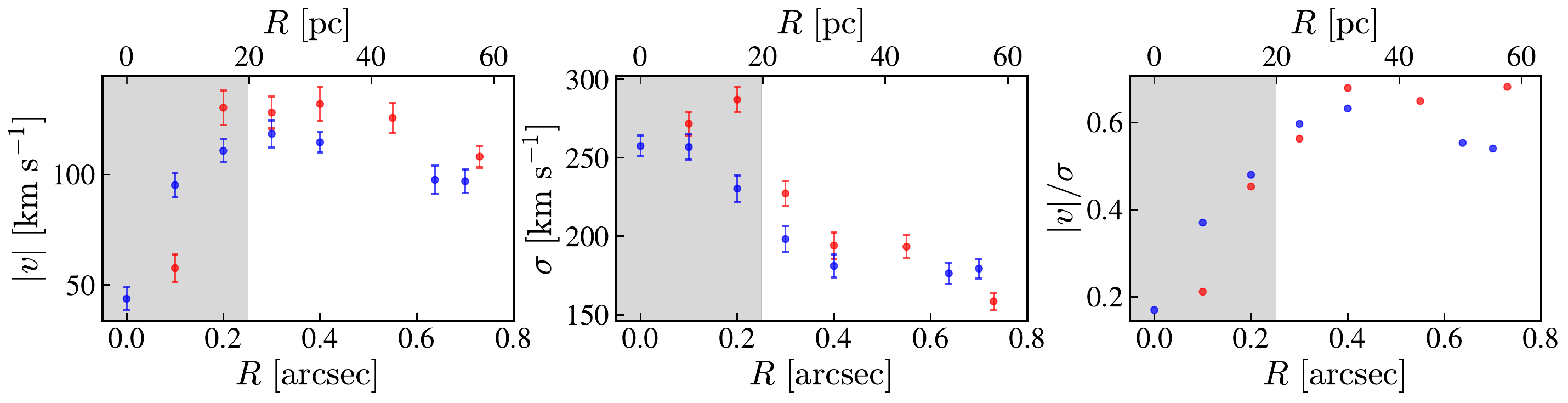}
	\caption{{\it First row, right panel:} Unsharp-masked WFPC2/F555W image of the double nucleus in NGC 4486B, showing two peaks separated by $\sim$12 pc. {\it First row, left and middle panels:} Stellar kinematic maps of $v$ and $\sigma$ from JWST-NIRSpec IFU data, shown in Voronoi-binned cells. In all panels, surface-brightness contours from the deconvolved WFPC2/F555W image are overplotted. The black dot marks the center of the faint peak (P2), which coincides with the $\sigma$ peak from NIRSpec and is identified as the likely location of the SMBH. The purple dot indicates the center of the bright peak (P1), coinciding with the center of the IFU data cube. The vertical dashed line marks the kinematic center. Second row: One-dimensional kinematic profiles of NGC 4486B derived from JWST-NIRSpec IFU data, showing the $|v|$, $\sigma$,  and $|v|/\sigma$ values along the major axis intersecting both nuclei. Red points represent measurements and associated uncertainties from the right side of the kinematic maps, where $\sigma$ peaks. Blue points correspond to the left side of the galaxy, mirrored about the rotation axis for direct comparison. The gray shaded region marks the approximate radial extent of the eccentric disk inferred from photometric and kinematic features (corresponds to the white rectangle above).}% \mv{I think it would help to add isophotal contours on the left plot. Also we should mark the kinematic center e.g. with a vertical line at $v=0$ to make clear it is not at $0\arcsec$.}
	\label{fig:phot}%
\end{figure*}

In Paper 1, we used the multi-Gaussian expansion (MGE) method \citep[]{Cappellari.2002} to model the surface brightness of NGC 4486B. The details of the fit are provided in Table 1 of Paper 1. In the top panel of Figure~\ref{fig:3d_profile}, we present the projected luminosity density of NGC 4486B (black dots), along with the corresponding models: the MGE model (in red) and S\'{e}rsic models with one (green), two (blue), three (orange), and four (gray) components. The vertical dashed line indicates the PSF FWHM of the F850LP image. None of the models with multiple S\'{e}rsic components are able to replicate the flatness of the inner profile as well as the MGE model does in this region. To test whether the presence of the double nucleus affects the derived surface brightness profile, we allowed the centers of the S\'{e}rsic (or Gaussian) components to vary freely in our GALFIT modeling. In the best-fit model, the centers of all components are found to lie within $\leq 1$ pixel of each other. Likewise, in the standard MGE fitting, where all Gaussian components share a fixed center at the brightest peak, the result remains consistent. The choice of fixing or freeing the component centers therefore does not significantly affect the shape of the final luminosity profile. This is likely because, although the nuclear region is slightly lopsided rather than fully symmetric, the two peaks are separated by only $\simeq$ 3-4 pixels, comparable to the PSF FWHM. Consequently, the central light distribution appears smooth rather than distinctly double in the F850LP image, causing all model components to converge to nearly the same photocenter, which lies on the bright peak.

%This is because the MGE model uses nested uniform density Gaussians to represent the surface brightness, resulting in a naturally flatter profile compared to the power-law S\'ersic models. 

For the MGE model we determined the 3D luminosity density of NGC 4486B from the projected luminosity density profiles by deprojecting the 2D MGE components into a 3D axisymmetric distribution, assuming an inclination angle of $\theta = 70^\circ$ (inferred from the Schwarzschild modeling in Paper 1). For the S\'ersic models, we numerically compute the 3D luminosity density by calculating the deprojection integral, assuming spherical geometry. The bottom panel of Figure \ref{fig:3d_profile} displays the deprojected 3D luminosity density profiles of each model along the semi-major axis. The inner luminosity profile of the MGE model is shallower than the multi-component S\'ersic models, which aligns with expectations from the 2D surface brightness profiles. 

The luminosity density profile in Figure~\ref{fig:3d_profile} (lower panel) shows that NGC~4486B  has a core that extends out to $\sim0.25\arcsec$ (20 parsecs). The luminosity density $\nu(r)$ in this region is described by $\nu(r) = r^{-\gamma}$  where the cusp slope  $\gamma = 0.14$.
%\mv{need to make sure this is correct - other values are mentioned else where 20 pc for the eccentric disk edge}

As mentioned in Section~\ref{sec:intro}, the favored mechanism for core formation in elliptical galaxies is  ``core scouring by an SMBH binary''. The low-density core produced in this manner can extend up to five times the black hole binary influence radius  \citep{rantala2018,Nasim2021,Khan2025}. For the black hole in NGC~4486B $\mbh=3.6\times10^8$\msun\ (Paper 1), the sphere of influence is estimated to be $20-35$~pc, which is comparable to the observed core radius. %\mv{check: ($\sim 13.5$pc ) is on the smaller end of the predicted range}.

The surface brightness fitting above ignored the double nucleus, which is barely resolved in the ACS/WFC F850LP image. In the higher spatial resolution of the HST WFC F555W image \citep{Lauer.1996} however, it is clearly resolved  (Figure 1 of Paper 1 gives both images). Figure~\ref{fig:phot}~(first row, right panel) shows an unsharp-masked image of NGC~4486B, in which the double nucleus is revealed with striking clarity. The two peaks are distinctly separated by $\sim$12 pc, with the fainter component (P2) marked by a black dot while the brighter nucleus (P1) is marked by a purple dot. Photometrically, P1 has a central surface brightness of $\mu_{I} = 12.89$ mag arcsec$^{-2}$, while P2 has $\mu_{I} = 13.00$ mag arcsec$^{-2}$, with a minimum of $\mu_I = 13.06$ mag arcsec$^{-2}$ between them \citep{Lauer.1996}. The white rectangle in this image encloses the nuclear region that we consider in greater detail below.

%\begin{figure*}
%	\centering	%
%	\includegraphics[width=2\columnwidth]{figure/Galfit_models.pdf}
%	\caption{The left panel presents the F850LP image, with a small blue square at the center indicating the field of view of the JWST IFU data cube. The second to fourth panels show the residual images from GALFIT when modeled with one, two, three, and four S\'{e}rsic functions, respectively.}%
%	\label{fig:galfit}%
%\end{figure*}

\subsection{JWST NIRSpec-IFU kinematics \label{sec:JWST}}

The spectroscopic data used in Paper 1 was obtained with the integral field unit mode of the JWST Near-Infrared Spectrograph \citep[JWST-NIRSpec IFU;][]{Boker.2022, Jakobsen.2022} as part of Cycle 1 Program 2567 (PI: M. Taylor). The observations were carried out using the G235H grating paired with the F170LP filter, covering the $1.66-3.17~\mu$m wavelength range.%and providing a resolving power that increases from $R \approx 2200$ at $1.66\,\mu$m to $R \approx 4400$ at $3.17 \mu$m. 
The NIRSpec IFU provides $3\arcsec \times 3\arcsec$ datacubes with $0.1\arcsec$ spaxels, corresponding to a spatial scale of $\sim8$ pc at the distance of NGC 4486B (for more details of the NIRSpec-IFU observations and data reduction see Paper 1)%The total exposure time was $\sim700$ seconds, obtained via a four-point dither pattern with NRSIRS2RAPID readout. A background field was observed with the same setup to allow accurate background subtraction. The final datacube was processed using the official JWST pipeline with additional custom routines to clean residual artifacts and improve calibration accuracy.

The kinematic datacube was Voronoi-tessellated \citep{Cappellari.2003_voronoi} to achieve a uniform signal/noise ratio of $\sim 80$ across the NIRSpec field-of-view.  Line-of-sight stellar kinematics were extracted using the penalized pixel-fitting method \citep[\texttt{pPXF};][]{Cappellari.2023}, fitting each Voronoi-binned spectrum with synthetic templates from the PHOENIX library \citep{Husser.2013}. We model the line-of-sight velocity distribution (LOSVD) using a Gauss-Hermite (GH) expansion as implemented in \texttt{pPXF} to extract mean velocity $v$, the width $\sigma$ of the best-fit Gaussian, and deviations captured by  $h_{3}$ (related to skewness) and $h_{4}$ (related to kurtosis) \citep[e.g.,][]{Marel.1993, Gerhard.1993}.  %While $\sigma$ characterizes the width of the underlying Gaussian rather than the true second moment of the LOSVD, we adopt ``velocity dispersion'' as a shorthand for $\sigma$ throughout this work for clarity.  

The left and middle panels of the first row in Figure~\ref{fig:phot} show $\vlos$ (in the galaxy rest-frame) and the velocity dispersion maps, respectively, obtained from JWST-NIRSpec.  The white rectangle encloses the same spatial region in all three panels.  The left-hand panel shows that the galaxy's central velocity (vertical dashed line) lies between P2 (black dot) and P1 (purple dot), but is closer to P1.  It is clear from the middle panel that the peak of the velocity dispersion (reddest bin) coincides with the fainter peak (black dot), which we therefore assume to be the position of the SMBH. The  $\sigma$ peak, measured from NIRSpec, has a maximum value of $\sigma = 287.2 \pm 7.4 \kms$. This offset in the velocity dispersion peak is also seen in the second row of Figure~\ref{fig:phot}, which shows 1D-kinematic profiles through the major axis of NGC~4486B. In this figure, the left side of the kinematic maps is mirror-reflected about the rotation axis to enable easier comparison of the two sides, and $|v|$ and $\sigma$ along the major axis are plotted. The vertical grey band shows the region encompassed by the white rectangle in Figure~\ref{fig:phot}. In addition to the fact that the velocity dispersion profile is not symmetric about the center, we see in the top panel that $|v|$  of stars in the vicinity of the presumed SMBH is higher than for stars on the opposite side of the galaxy by $16\kms$ The rightmost panel shows the ratio  $|v|/\sigma$, indicating that this galaxy has roughly equal amounts of rotational and dispersion support over the range of radii examined. %\mv{need to recheck value when mirroring is done wrt rotation axis not location of P1}.

%\textbf{At first glance, the increase of $|v|$ with radius and the nearly flat $\sigma$ profile within the two central spatial bins around the brightest nuclear peak may appear counterintuitive when compared to expectations for a cold, circular Keplerian disk. However, END are dynamically hot systems composed of apsidally aligned, eccentric stellar orbits rather than circular ones. In such systems, stars spend a significant fraction of their orbital time near apoapsis, which suppresses the net line-of-sight streaming velocity close to the nucleus, while the superposition of radial motions and mixed orbital phases maintains a high velocity dispersion. As a result, the innermost region of an END can appear less rotationally supported (lower $|v|/\sigma$) than regions at slightly larger radii, where streaming motions become more coherent. This behavior is consistent with theoretical and numerical studies of eccentric nuclear disks and with observations of the M31 nucleus \citep{Tremaine.1995, Peiris.2003}.}

Taken together, the resolved double nucleus, the offset of the velocity-dispersion peak from the photometric maximum, and the asymmetric major-axis $|v|$ and $\sigma$ profiles shown in Figure 2 provide purely observational photometric and kinematic signatures that are most naturally explained by an END in NGC~4486B.

\subsection{Schwarzschild model and $N$-body model for NGC 4486B \label{dynamical_model}}

In Paper 1, we combined the photometric data described in Section~\ref{sec:photometry}, the JWST kinematics summarized in Section~\ref{sec:JWST}, and supplementary long-slit CFHT/SIS measurements from \citet{Kormendy.1997} to construct dynamical models of NGC~4486B and constrain both the black hole mass and the large-scale mass distribution.

The two-dimensional JWST kinematic maps and the unsharp-masked image shown in Figure~\ref{fig:phot}—presented here for completeness—reveal a well-resolved double nucleus and a velocity-dispersion peak offset from the photometric center. In Paper 1, we applied two independent dynamical approaches to these data: axisymmetric Schwarzschild orbit-superposition modeling \citep[e.g.][]{Schwarzschild.1979,vanderMarel.1998,Cretton.1999,Gebhardt.2003,Valluri.2004,Thomas.2004,vandenBosch.2010,Vasiliev.2020A}, which fits the full LOSVD including higher-order moments, and axisymmetric Jeans Anisotropic Modeling \citep[JAM;][]{Cappellari.2008}, which uses only the first two velocity moments. Both approaches yielded broadly consistent SMBH mass estimates. We refer the reader to Paper 1 for full details of these models and results.

In the present work, we restrict our discussion to the orbit-superposition models constructed with the FORSTAND code \citep{Vasiliev.2020A}, using them to examine the orbital structure and implications for the formation of the eccentric nuclear disk.

As both  JAM and Schwarzschild modeling assume global symmetry and steady-state equilibrium, they cannot reproduce the pronounced asymmetries in the JWST-NIRSpec IFU kinematics and light distribution associated with the double nucleus. To gauge the impact of this modeling limitation, in Paper 1 we analyze three data configurations: (i) the original asymmetric kinematic maps; (ii) maps in which spaxels at and around the peak of the velocity dispersion, $\sigma$, are masked across all moments, suppressing the strongest asymmetry allowing us to probe the minimum black hole mass allowed; and (iii) a final test in which only the $\sigma$ map was modified by shifting its peak to the photometric center,  an intentionally unphysical but informative symmetrization used to explore the maximum mass permitted under the symmetry assumption. These models yielded $2.8^{+0.6}_{-0.4}\times 10^8 \leq \mbh/\msun \leq5.3^{+0.9}_{-0.9}\times 10^8$, the detection significant at $>3\sigma$ in all models. The inclusion or exclusion of a dark matter halo has a negligible impact on the measured $\mbh$. Our Schwarzschild models, fit directly to the original JWST kinematics (no shifting or masking of the peak), yielded a black hole mass $\mbh = 3.6\pm{0.7} \times 10^8$\msun, a stellar mass-to-light ratio of $M/L_{[F850LP]} = 3.9\pm0.5$ corresponding to a stellar mass $M_*=7.0\pm0.8\times 10^9\msun$, and an inclination $\theta\sim 70^\circ$.   

In addition to providing the globally optimal model parameters, the Schwarzschild method delivers the full orbital distribution function through the weights assigned to each orbit in the library. While the black hole mass and stellar mass reported above represent marginalized estimates (median and uncertainties) over the full model ensemble, the construction of an $N$-body realization requires selecting a single best-fitting model corresponding to the minimum $\chi^2$. These weights allow us to generate a self-consistent, data-driven $N$-body realization of the galaxy.

Although one of our best-fitting models includes a modest dark matter fraction ($\sim14\%$ by mass), for the purposes of this study, we construct the $N$-body model using only the stellar component in order to simplify the subsequent dynamical experiments. The initial conditions for our simulations are drawn from the minimum-$\chi^2$ Schwarzschild model, with $\mbh=3.51\times10^8 \msun$, mass-to-light ratio $\Upsilon_{[F850LP]}=4.1$, a total stellar mass $M_*=7.44\times10^9\msun$, and an intrinsic axisymmetric shape with axis ratio $0.83$. These values differ slightly from the marginalized medians but lie well within the quoted uncertainties.

We emphasize that this $N$-body realization is based on the axisymmetric Schwarzschild model and therefore does not explicitly contain an END. The purpose of these simulations is solely to trace the large-scale trajectory and damping timescale of a recoiling SMBH in the global potential of NGC~4486B. They are not intended to model the formation or internal structure of the END. A fully self-consistent, non-axisymmetric $N$-body treatment tailored to the observed eccentric disk structure is beyond the scope of the present work and will be explored in future studies.

\section{Formation of an Eccentric Nuclear Disk from a Recoiling SMBH}
 \label{sec:END_recoil}

\subsection{Comparison of nuclear structure and kinematics of  NGC~4486B with eccentric nuclear disk simulations \label{sec:ENDsims_VS_Schw}}

\citet{Wernke.2021} used $N$-body simulations to produce mock photometric and two-dimensional kinematic maps of ENDs for a wide range of viewing orientations. Their results show that the observed morphology depends strongly on the angle between the eccentricity vector (the disk’s major axis) and the line of sight. When the disk is viewed edge-on with its major axis aligned along the line of sight, only a single, centered nucleus is seen. If the disk is viewed edge-on with its minor axis along the line of sight, the nucleus appears offset. For edge-on or nearly edge-on views at intermediate angles relative to the eccentricity vector, the system appears as a double nucleus, where the fainter secondary peak lies close to the location of the black hole \citep[see Figure 2 in][]{Wernke.2021}. They also found that the velocity dispersion is generally elevated and peaks at the location of the SMBH, irrespective of the viewing angle, and this dispersion peak typically does not coincide with the photometric center (brightest peak) unless the END is viewed such that it appears as a single centered nucleus. 

We therefore expect a double nucleus in NGC 4486B, consistent with our nearly edge-on viewing angle ($\theta \sim 70^{\circ}$), inferred from our Schwarzschild models. The morphology in the right-hand panel of Figure~\ref{fig:phot} closely resembles the mock ENDs viewed nearly edge-on at intermediate angles to the eccentricity vector in \citet{Wernke.2021}, supporting this interpretation and consistent with the earlier conclusion of \citet{Lauer.1996}. In such nearly edge-on ($\theta \sim 70^{\circ}$) configurations, the velocity-dispersion peak is expected to lie near the SMBH rather than at the brightest photometric peak—a behavior clearly seen in Figure~\ref{fig:phot} (first row, middle). A further examination of the 1D kinematic profiles of NGC 4486B along the major axis intersecting both nuclei (the second row of Figure~\ref{fig:phot}) shows a clear asymmetry between the left and right sides of the galaxy: $|v|$ is elevated by $\sim16\kms$ and $\sigma$ is elevated by $\sim57\kms$ compared to the opposite side of the galaxy. Similar kinematic asymmetries and the offset of the velocity dispersion peak from the brighter peak strongly support the eccentric disk simulations \citep{Wernke.2021}, which also suggests local dynamical disturbances near the velocity dispersion peak  (the presumed location of the SMBH). The enhanced $|v|$  may reflect streaming motions within the eccentric disk or, alternatively, residual motion of the SMBH relative to the center of the galaxy. Since the mock kinematics and images in  \citet{Wernke.2021}  were produced from simulations of isolated ENDs surrounding SMBH, whereas the one in NGC~4486B is embedded in an elliptical galaxy, additional simulations are needed to fully understand the offsets.

 \begin{figure}
	\centering	%
	\includegraphics[width=\columnwidth]{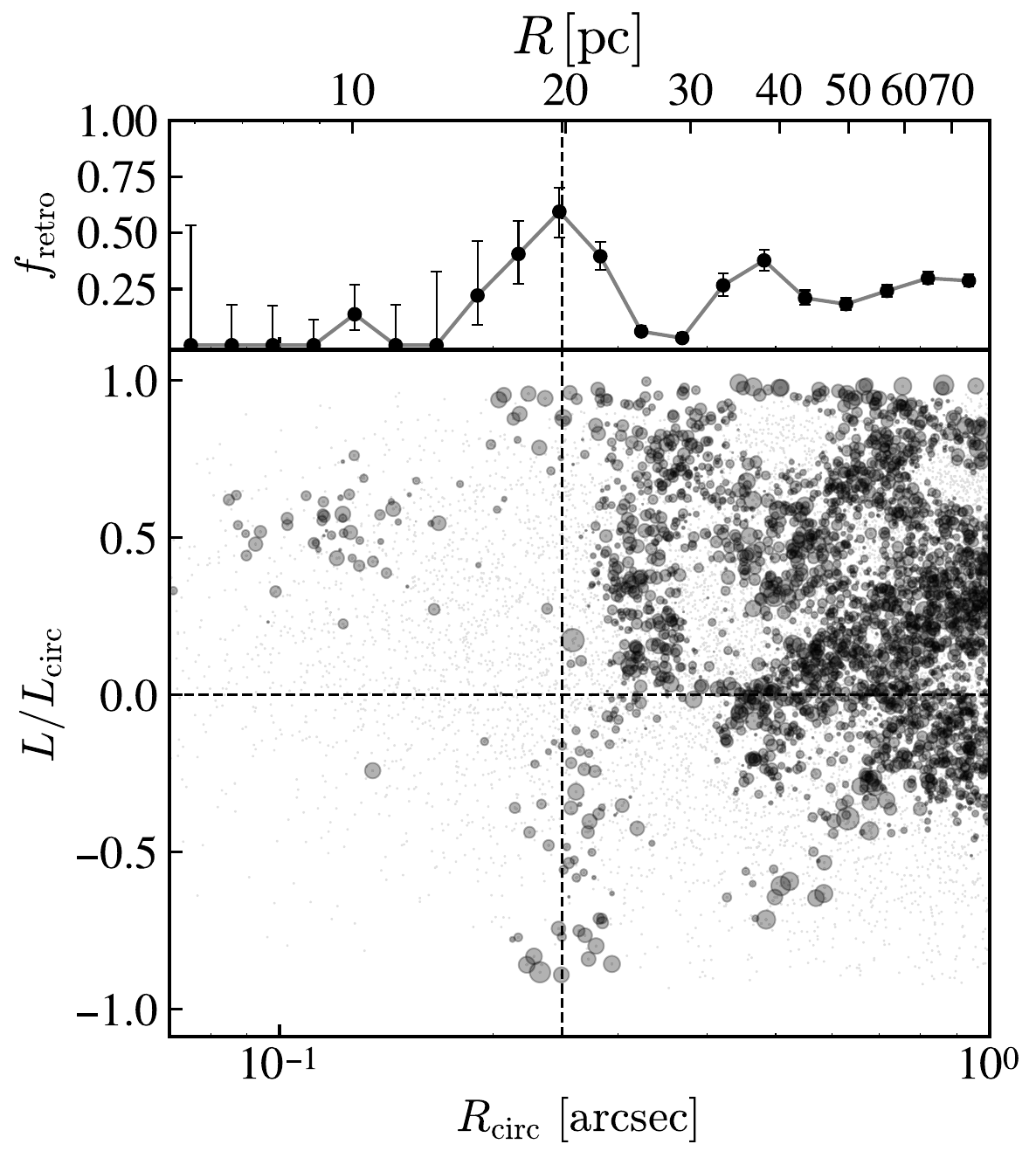}
\caption{Top: Radial profile of the retrograde-orbit fraction from the best-fit Schwarzschild model using the high-resolution orbit library (100,000 orbits). The vertical dashed line at $0.25\arcsec \simeq 20$ pc marks the boundary of the eccentric-disk region highlighted by the white rectangle in Fig.~\ref{fig:phot}. Bottom: Distribution of orbits in phase space, showing mean radius versus orbital circularity. The size of each circular marker scales with the relative weight of the corresponding orbit.}
	\label{fig:ret_fract}%
\end{figure}

 %\begin{figure*}
%	\centering	%
	%\includegraphics[width=0.75\textwidth]{figure/modifiedFigs3a.png}
%\caption{Top-left: unsharp masked region within white rectangle in Fig.~\ref{fig:phot}. Top-right: Radial profile of the retrograde-orbit fraction for the best-fit Schwarzschild model fitted to the original kinematic dataset with a large orbit library (100,000 orbits). The vertical dashed line is at $0.25\arcsec \simeq 20$pc and marks the edge of the eccentric disk region highlighted by the white rectangle in Fig.~\ref{fig:phot}.
%Bottom-left: 2D-histogram of simulated END inclined at $70^\circ$ showing black hole (black dot) \mv{hopefully they sort of resemble each other}.
%Bottom-right: retrograde orbit fraction in simulated eccentric disk as a function of radius (from BH?)}
%	\label{fig:END-retfract}%
%\end{figure*}

%\mv{describe comparison between Fig~\ref{fig:END-retfract}~top-left and bottom-left when figs are finalized.}

As mentioned in Section~\ref{sec:intro}, one of the proposed formation pathways for ENDs is a GW recoil of an SMBH following coalescence \citep{Akiba.2021, Bright.2024}. Such a recoil can drive stars that were initially on nearly circular prograde orbits onto eccentric trajectories and flip a substantial fraction, particularly at larger radii, into retrograde motion, as the displaced SMBH effectively outruns the outer stars and reverses their angular momenta in its rest frame.

The resulting disk therefore contains a mixture of prograde and retrograde orbits, with the retrograde fraction increasing with radius; more powerful kicks generate a correspondingly larger retrograde population \citep{Bright.2024}. Simulations further indicate that this transformation occurs on very short timescales - within only a few dynamical times, $\sim10^{4}-10^{5}$ yr on parsec scales \citep{Akiba.2021}.  In cases where the disk hosts a substantial retrograde component, apsidal precession slows dramatically, allowing the lopsided eccentric disk structure to survive for $\gtrsim10^{8} -10^{9}$ yr as a long-lived dynamical fossil of the recoil  \citep{Madigan.2018}.

Before proceeding, we emphasize that the analysis presented below is not intended to establish the existence of the END itself. The observational evidence for an END in NGC~4486B is provided directly by the photometric and kinematic signatures discussed in Section~\ref{sec:data} and shown in Figure~\ref{fig:phot}. Instead, the purpose of the Schwarzschild-based analysis presented here is to examine the orbital phase-space structure required by the observed kinematics and to assess whether it is consistent with the GW recoil kick-based END formation scenarios.

Although our Schwarzschild dynamical models from Paper 1 do not explicitly model the double nucleus, the fact that they were designed to fit the observed kinematics provides a window into the orbital architecture of the nuclear region. We now examine the best-fit Schwarzschild model from Paper 1 to assess how the fraction of retrograde orbits varies with radius. For each orbit we extract mean radius \(R\), and orbital circularity \(\lambda = L_z/L_{\rm circ}\) (retrograde if \(\lambda<0\)), $L_z$ is the angular momentum of the orbit and $L_{\rm circ}$  the angular momentum of a circular orbit with the same binding energy.  We keep the orbits with \(0 \le R \le 1^{\prime\prime}\), then divide this range into 20 uniform bins.
Since the orbit weights ($w$) in the Schwarzschild model correspond to the mass of stars on a given orbit, in each bin we compute the weighted share of retrograde orbits (retrograde weight divided by total weight in that bin). We compute uncertainties that reflect both the retrograde fraction and the number of effectively independent orbits contributing to a bin, reporting central 68\% credible intervals using a Jeffreys prior binomial model \citep{Jeffreys.1946} with an effective sample size derived from the orbit weights. To test robustness against discreteness noise and finite library size, we repeated the modeling in Paper 1 and measurement of the retrograde orbit fraction with an orbit library five times larger  (100{,}000 orbits) than our standard model in that paper. The resulting profiles from the original and large-library model agree within the quoted uncertainties, indicating that our conclusions are not driven by sampling effects. We only show the results for the large-library model. The top panel of Figure~\ref{fig:ret_fract} shows the retrograde-orbit fraction as a function of radius for the Schwarzschild model, while the bottom panel shows the corresponding orbital circularity distribution.  The vertical dashed line at \(R=0.25^{\prime\prime}\) (\(\sim 20\,\mathrm{pc}\)) marks the edge of the eccentric disk region in the white rectangle in Figure~\ref{fig:phot}. Our results closely resemble the orbital distribution of an eccentric nuclear disk formed via a recoil kick \citep{Akiba.2023}, which also exhibits a substantial retrograde population ($\sim$50\%), particularly in the outer regions of the disk. In Figure~14 of \citet{Akiba.2023}, the top panel shows the orbital circularity distribution and the bottom panel shows the retrograde-orbit fraction. Although our Schwarzschild models assume axisymmetry and therefore cannot reproduce the full asymmetric structure of an eccentric disk, they robustly demonstrate that the observed JWST kinematics require a substantial retrograde orbital component. This comparison is therefore used to assess consistency with proposed END formation scenarios, rather than to establish the existence of the END itself.

%Before proceeding, we emphasize that the analysis presented below and summarized in Figure~\ref{fig:ret_fract} is not intended to establish the existence of the END itself. The observational evidence for an END in NGC~4486B is provided directly by the photometric and kinematic signatures discussed in Section~\ref{sec:data}. Instead, the purpose of the Schwarzschild-based analysis presented here is to examine the orbital phase-space structure required by the observed kinematics and to assess whether it is consistent with specific END formation scenarios.

\citet{Cretton.1999} demonstrated that Schwarzschild models can sometimes assign artificially large weights to retrograde orbits when the kinematics are constrained using only the GH expansion of the LOSVD. In their study of a rapidly rotating S0 galaxy, models without a central black hole reproduced the measured $h_3$ and $h_4$ moments by populating a large fraction of retrograde (counter-rotating) orbits, even though the observed velocity profiles showed no such component. 
 This ``spurious counter-rotation'' arises because truncated GH series cannot uniquely describe the full LOSVD, allowing the model to exploit retrograde orbits as a fitting freedom. Importantly, this effect is most pronounced in cold, rotationally supported systems with $|v|/\sigma \gtrsim 1.5$, where small distortions in the LOSVD skewness can be mimicked by a retrograde tail. 

By contrast, NGC 4486B is a much more dispersion-dominated system: as shown in the bottom right panel of Figure~\ref{fig:phot}, we measure $|v|/\sigma \lesssim 0.45$ in the inner regions, increasing only to $\sim 0.6$ at the outer edge of our JWST-NIRSpec field of view. In this regime, the degeneracies highlighted by \citet{Cretton.1999} are much less severe, making it unlikely that the strong retrograde signal we find is purely an artifact of the GH parameterization. Instead, the retrograde component appears to be an intrinsic dynamical requirement of the system.

To test the importance of retrograde orbits explicitly, we re-ran our best-fit Schwarzschild models with identical initial conditions but excluded all retrograde orbits from the library. This increased the $\chi^2$ of the best-fit model by $\sim 23\%$, and the $\sigma$ map becomes vertically elongated and flattened along the projected minor axis. In contrast, the model including retrograde orbits produces a more rounded $\sigma$ structure, in much closer agreement with the observed data.  Figure~\ref{fig:Schw_Ap} in the Appendix compares the resulting two-dimensional $v$ and $\sigma$ maps: the first row shows the best-fit model with retrograde orbits, while the second row shows the best-fit model with all retrograde orbits removed. This confirms that the retrograde component is required to reproduce the kinematics of NGC~4486B. The presence of a significant retrograde orbit fraction in the region that appears to be dominated by the END, thus offers tantalizing evidence that the observed END is consistent with formation via a past SMBH recoil event.

In addition to the recoiling black hole scenario, simulations show that the hardening of an SMBH binary can itself produce a large retrograde fraction in the nuclear orbital distribution: three-body slingshot interactions preferentially deplete low-$L$ and co-rotating (prograde) stars, leaving a tangentially biased remnant in which stable, counter-rotating orbits dominate inside the binary's sphere of influence \citep[e.g.,][]{Meiron.2010, Rantala.2019}. %This can form a kinematically decoupled, often counter-spinning core \citep[e.g.,][]{Meiron.2010, Rantala.2019}. 
Given that NGC 4486B hosts both a flat stellar core and a central SMBH, a binary-scouring scenario provides an alternative explanation for the large retrograde fraction we infer from the Schwarzschild model, offering complementary support for the possibility that the black hole in NGC 4486B is the result of the merger of a binary black hole. 
%\mv{do we need to say something about why the retrograde fraction rises again in the outer part of the galaxy?} \BT{I would suggest adding a few sentences outlining a possible hypothesis by TA, but phrased as a tentative interpretation rather than a firm conclusion, or maybe not!}

\subsection{Constraints on the Recoil Kick and SMBH Binary Parameters}\label{sec:kick_q_estimate}
\begin{figure}[t!]
\centering
\includegraphics[width=\linewidth]{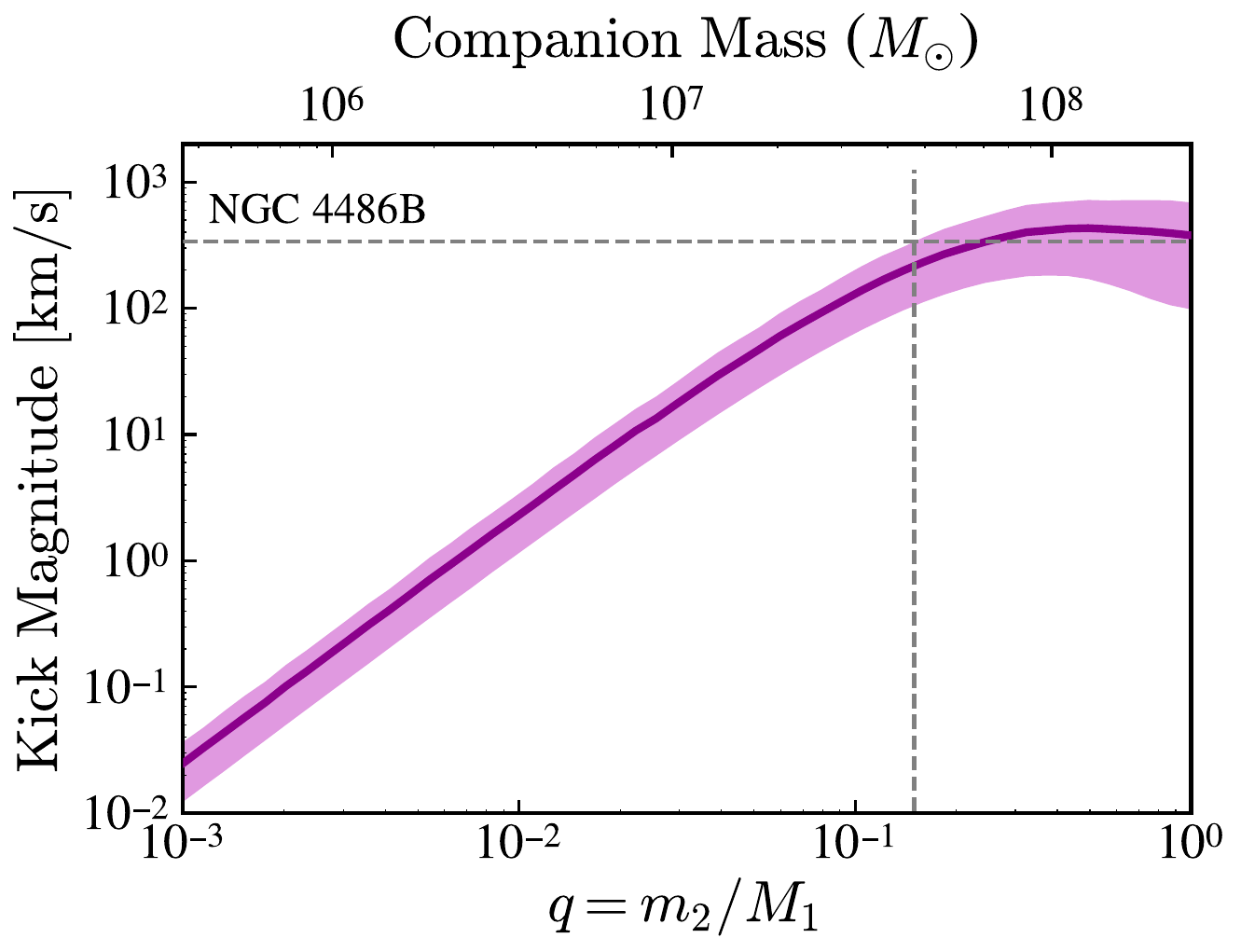}
\caption{Expected GW recoil kick magnitude, $V_{\rm{kick}}$, as a function of the pre-merger black hole mass ratio, $q$. The thick purple line shows the mean, and the shaded region a $1 \sigma$ confidence interval estimated via Monte Carlo simulations. The recoil kick required to induce the double nucleus of NGC 4486B is shown in the gray horizontal dashed line, and the mass of the corresponding pre-merger companion to NGC 4486B's central SMBH is indicated on the top horizontal axis. The gray vertical dashed line marks $q = 0.15$, the lower-limit  pre-merger black hole mass ratio for NGC 4486B. Adapted from \citet{Akiba.2025} which uses the analytical prescription from \citet{Lousto.2010, Lousto.2012}.}
\label{fig:recoil_kick_mag}
\end{figure}

\citet{Akiba.2021} simulated disks with stars initially in circular orbits about an SMBH. Following an in-plane GW recoil kick imparted on the black hole, the stars preferentially align their eccentricity vectors perpendicular to the recoil kick with magnitude
\begin{equation}
|\vec{e}_{\rm{avg}}| = \frac{3}{2} \,
 \frac{V_{\rm{kick}}}{v_c} \, ,
\label{eqn:e_avg}
\end{equation}
where $V_{\rm{kick}}$ is the recoil kick speed and $v_c$ is the initial average circular speed of the stars. 
In their simulations, they found that immediately after the kick, the mean eccentricity jumps to $|\vec{e}_{\rm{avg}}| \sim 1$ but subsequently decreases via large amplitude oscillations to smaller values. 
Simulations show that while the disk gradually diffuses in semi-major axis and orientation (as $|\vec{e}_{\rm{avg}}|$ decreases), $R_c$, the characteristic radius of the disk and the degree of apsidal alignment remain stable over $\gtrsim10$ Myr \citep{Akiba.2021, Akiba.2023}. 
\citet{Akiba.2023} noticed that by setting $|\vec{e}_{\rm{avg}}| \, \sim 1$, one can determine $R_c$ of the resulting eccentric disk as,
\begin{equation}
R_c \equiv \frac{4}{9} \, \frac{G M_{\rm BH}}{V_{\rm{kick}}^2} \, ,
\label{eqn:r_c}
\end{equation}
where $M_{\rm BH}$ is the mass of the recoiling black hole. This $R_c$ is where we expect the eccentricity and apsidal alignment of the post-kick disk to be maximized.  Applying these analytics to NGC 4486B with $M_{\rm BH} = 3.6 \times 10^8 \, M_{\odot}$ and $R_c \sim 6$ pc (taking the double nucleus separation of 12 pc to be the distance between the periapsis and apoapsis of the eccentric disk), we estimate (using eq.~\ref{eqn:r_c}), a required recoil kick $V_{\rm{kick}} \approx 340$\kms. This is a fairly moderate magnitude for a GW recoil kick (see Figure \ref{fig:recoil_kick_mag}).

In Figure \ref{fig:recoil_kick_mag}, we show the expected recoil kick magnitude as a function of the pre-merger black hole mass ratio, $q=m_2/M_1$ (where $M_1$ is the more massive black hole in the binary). The kick magnitudes are computed using analytic fitting formulae calibrated to numerical relativity simulations of binary black hole mergers \citep{Lousto.2010, Lousto.2012}, as implemented in the Monte Carlo framework of \citet{Fragione.2023} and applied to END scenarios by \citet{Akiba.2025}. The light purple region shows the Monte Carlo results of $10^6$ evaluations of the analytical model (described in \cite{Fragione.2023}) with randomized spins. The dimensionless spins for both black holes are drawn from an isotropic distribution in direction and a uniform distribution in magnitude ([0, 1]). The thick purple line shows the average kick magnitude as a function of $q$ marginalized over the spin distribution. The mean and $1 \sigma$-confidence interval of the recoil kick distribution is shown, and the estimated $V_\mathrm{kick}\sim 340\kms$ , required to induce NGC 4486B's double nucleus is indicated with the dashed gray line. This recoil kick implies a pre-merger mass ratio of $q \approx 0.24$ which corresponds to a predicted companion mass of $7 \times 10^7 \, M_{\odot}$ (the calculation does not account for mass lost to GWs). 
However, the estimated recoil kick magnitude falls within one standard deviation of the mean for all mass ratios $q > 0.15$, so one can only determine a \textit{lower limit} at the companion mass of $4.7 \times 10^7 \, M_{\odot}$. Additionally, it is important to note that these recoil kick and mass ratio estimates assume that stars are initially on circular orbits. \citet{Akiba.2023} showed that the inclusion of initial eccentricities reduces the kick magnitude required to produce an eccentric disk at a given characteristic radius, so a degeneracy exists between the companion mass estimate and the pre-merger stellar eccentricities in the nuclear star cluster. This analytic framework establishes the connection between the eccentric disk geometry and the GW recoil kick.  
 
We emphasize that this analytic treatment is intended to provide order-of-magnitude constraints and a consistency check on the recoil scenario, rather than a fully predictive, galaxy-specific model. A direct comparison between theory and the detailed observables of NGC~4486B—such as generating surface-brightness and two-dimensional kinematic maps from tailored $N$-body simulations—is beyond the scope of the present paper. Existing END simulations are intentionally idealized, typically modeling isolated eccentric disks with limited particle numbers to capture the underlying physical mechanism and viewing-angle dependencies, whereas a realistic forward model for NGC~4486B would require high-resolution simulations that self-consistently include both the stellar bulge and the nuclear disk and explore a broad parameter space (disk structure, eccentricity distribution, kick amplitude and direction, and viewing angle). We therefore focus here on establishing plausibility and placing data-driven constraints on the recoil kick scale using analytic scalings and the observed kinematic signatures, and view galaxy-specific modeling as a natural next step to be pursued in future work.
 
We build upon the above analysis in the following sections to explore the plausible timescales of such an event and the dynamical signatures it may have imprinted on the double nucleus in NGC 4486B.

\begin{figure*}
    \centering
    \includegraphics[width=\linewidth]{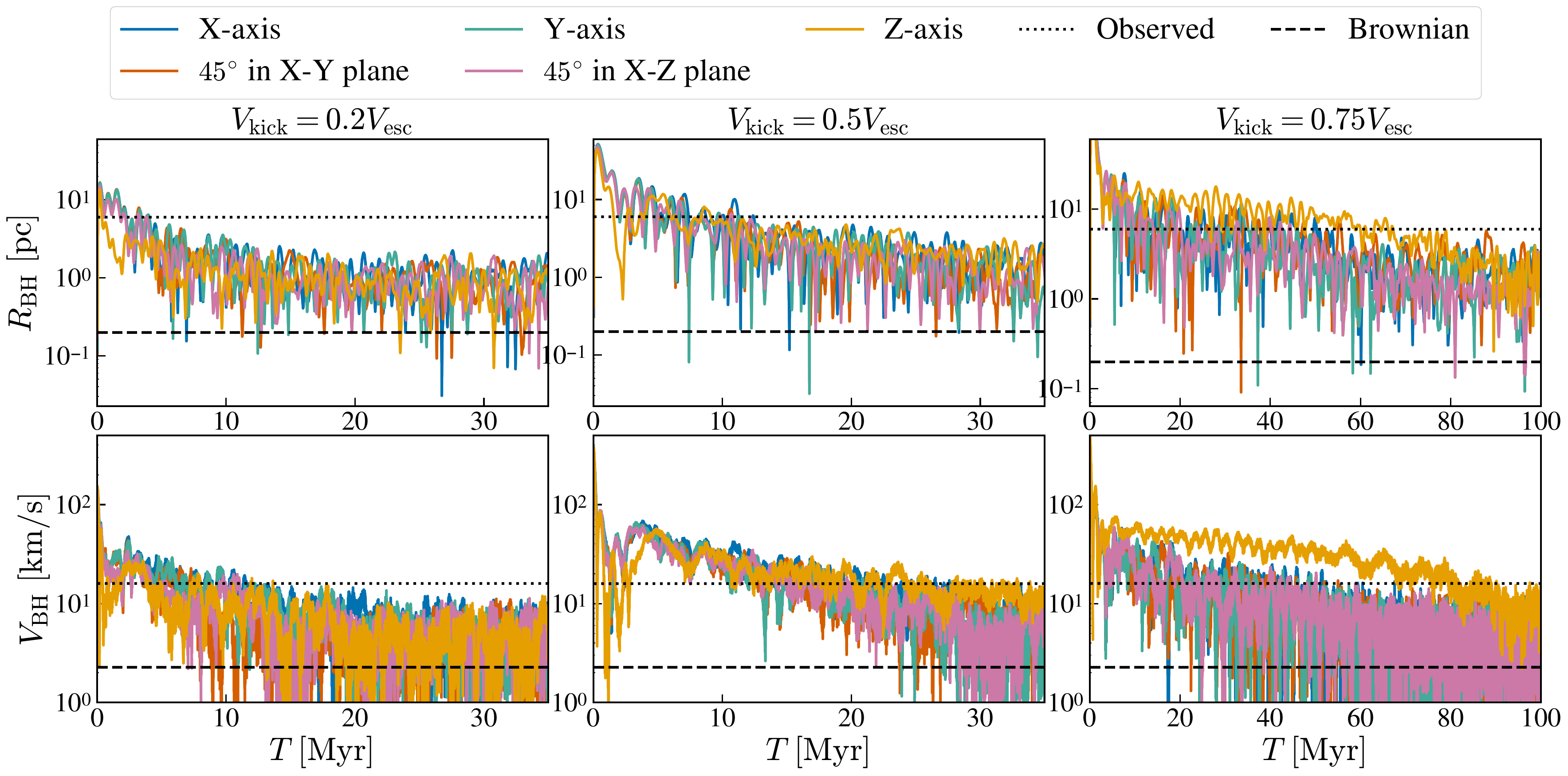}
    \caption{Trajectories of the black hole that is kicked from the center of NGC~4486B. The different colored curves represent kicks in different directions. The top panels show the BH's relative distance from the galaxy center, and the bottom panels show its relative speed. The left and right panels correspond to a kick velocity of $V_{\rm kick}=0.2 V_{\rm esc}$, $0.5 V_{\rm esc}$, and $0.75 V_{\rm esc}$, respectively. The black dashed and dotted lines show the expected amplitude of Brownian motion and the observed offset in position and $\vlos$, respectively. In all cases, the black hole quickly returns close to the galaxy center, with position and velocity offsets that are well below the observed values.}
    \label{fig:trajectories}
\end{figure*}

\section{Post-Kick Trajectory of the Recoiling SMBH}
\label{ssec:nbody_kick}
Considering the strong evidence in support of a recoil kick, we now explore the possible post-kick trajectories of the SMBH. We construct an $N$-body representation of the best-fit Schwarzschild model of NGC~4486B with $10^6$ particles. This model, which has no dark matter halo, has a best-fit black hole mass of $M_{\rm BH}=3.51 \times 10^8\msun$, total stellar mass $M_*=7.44 \times 10^9\msun$, and mass-to-light ratio $\Upsilon_{[F850LP]}=4.1$. We introduce a black hole particle at the center of the system and give it a kick velocity $V_{\rm kick}$ along several directions. We explore $V_{\rm kick}=0.2 V_{\rm esc}=154\kms$, $V_{\rm kick}=0.5 V_{\rm esc}=384\kms$, and $V_{\rm kick}=0.75 V_{\rm esc}=576\kms$, where $V_{\rm esc}$ is the escape velocity from the galaxy center (the middle value being close to the kick velocity estimated in Section~\ref{sec:kick_q_estimate}). 

The simulations are run using the publicly available code \texttt{GyrfalcOn} \citep{Dehnen2002}. Each particle is softened using the P1 softening kernel\footnote{A modification of the Plummer softening kernel whose density is $\propto (1+x^2)^{-7/2}$}, with softening length 0.79 pc for the stars and 0.395 pc for the BH. We use hierarchical timesteps, where the timestep of each particle depends on its softening length $\epsilon$ and instantaneous acceleration $a_i$ as $\Delta t \sim 0.025\sqrt{\epsilon/\left| a_i \right|}$. According to this criterion, a circular orbit around the black hole at 0.79 pc (the softening length) is resolved with at least $\sim$100 timesteps. 

Figure \ref{fig:trajectories} shows properties of the resulting trajectory of the SMBH in the various cases. The top and bottom panels show the SMBH's distance and velocity with respect to the galaxy's center of mass\footnote{In order to calculate the center of mass, we use the $90\%$ most bound stars in the simulation, as some stars may get unbound during the recoil kick. The results differ only slightly based on the exact center of mass definition.}, and the right, middle, and left panels correspond to cases with $V_{\rm kick} = 0.2 V_{\rm esc}$, $0.5 V_{\rm esc}$, and $0.75 V_{\rm esc}$, respectively. Notably, in all simulations, the line-of-sight velocity dispersion peak consistently follows the SMBH, justifying our assumption in Section~\ref{sec:ENDsims_VS_Schw} that the off-center velocity dispersion peak marks the location of the SMBH. Within each panel, the various curves correspond to different initial kick directions. In all panels, the dotted black line marks the observed positional offset of the SMBH (6 pc) and the localized $\vlos$ offset (16 km s$^{-1}$) measured at the location of the SMBH (see top panel of Fig.~\ref{fig:phot}).  This feature is assumed to trace the mean motions of stars bound to the SMBH but may not represent the intrinsic 3D velocity of the black hole itself. However, for comparison with the simulations, we refer to this quantity as the ``observed line-of-sight velocity''. The black dashed lines in the lower and upper panels show the expected velocity and position amplitudes due to Brownian motion, respectively, which are given by \citet{Gualandris.2008}:
\begin{equation}
    V_{\rm Brown} = \sigma_0\sqrt{\frac{3 m_*}{M_{\rm BH}}} \quad \quad R_{\rm Brown} = \frac{V_{\rm Brown}}{\sqrt{4 \pi G \rho_c/3}}
\end{equation}
where $\sigma_0$ is the central velocity dispersion, $m_*$ is the mass of each star particle, and $\rho_c$ is the central density. 

In every trajectory shown in Figure~\ref{fig:trajectories}, the black hole rapidly sinks back toward the galaxy center, reaching amplitudes comparable to the observed positional and $\vlos$ offsets, after which these quantities fall below the observed values. The return timescale increases with kick strength, from $\sim$10 Myr for $0.2 V_{\rm esc}$ to $\sim$30 Myr for $0.5 V_{\rm esc}$ and $\sim$80 Myr for $0.75 V_{\rm esc}$. Although these timescales differ, all are extremely short relative to a Hubble time. Furthermore, the observed $\vlos$ offset is only one component of the possible 3D velocity of the SMBH. At late times, the black hole retains a residual velocity relative to the galaxy center that is clearly resolved above the expected Brownian amplitude. This residual motion is likely the result of a core-stalling effect \citep[e.g.,][]{Read2006}, but it remains insufficient to account for the observed velocity offset. Interestingly, the timescale for the black hole to return to the center is longer if the kick is along the Z-axis, compared to the other directions, although this timescale is still significantly shorter than a Hubble time. This rapid damping of the SMBH motion is consistent with recent work showing that the long-term trajectories of massive perturbers in galaxy cores depend sensitively on the core structure and are generally damped on relatively short timescales in realistic potentials \citep{Rawlings.2025}. This will be further investigated in a follow-up paper (Jha et al, in preparation). 

Taken together with our estimate of $V_{\rm kick}\sim 340\kms$ (Section~\ref{sec:kick_q_estimate}) ($\lesssim 0.5 V_{\rm esc}$), our results indicate that a black hole displaced from the galaxy center by a recoil kick will return within  $\sim 3\times 10^7$~yr, regardless of the kick’s direction or magnitude. This timescale is significantly shorter than the survival time of an END, which can remain stable for $\gtrsim 10^{8}$–$10^{9}$~yr \citep{Madigan.2018}, suggesting that the double nucleus and associated kinematic asymmetries in NGC~4486B are best explained as relics of a recent recoil event. A long-lived oscillatory ``sloshing'' of the SMBH through the galaxy \citep[e.g.,][]{Gualandris.2008} therefore appears highly improbable, probably because the SMBH in this galaxy is $\sim 10$\% of the galaxy's mass \mdash nearly two orders of magnitude larger than the typical black hole mass fraction of 0.1\% expected from SMBH scaling relations. 

It must be noted that the initial galaxy models we used for the simulations are based on the {\em current observations} of NGC~4486B. As discussed previously, oscillations of the SMBH following a recoil kick can scatter stars and flatten the central cusp beyond what would have resulted from ``binary black hole scouring'' alone. The density profile of NGC~4486B prior to the kick was therefore likely steeper (denser) than assumed in our simulations. A denser central region would enhance dynamical friction and further shorten the return timescale, implying that the black hole return times inferred from the trajectories in Figure~\ref{fig:trajectories} likely represent upper limits. In a forthcoming study, we will extend this analysis by modeling recoil trajectories in more strongly cusped pre-kick galaxy profiles.

%It must be noted that the initial galaxy models we used for the simulations are based on the {\em current observations} of NGC~4486B. As discussed previously, the oscillations of the SMBH following the kick scatter stars and flatten the cusp beyond what may have resulted from ``binary black hole scouring''. Hence the density profile of NGC~4486B prior to the kick was likely steeper (denser) than assumed in our simulations. However, a denser central region would only provide greater dynamical friction, further reducing the timescales. Hence, our simulations in Figure~\ref{fig:trajectories} may actually overestimate the timescale for the SMBH to return to the center.

In the following section, we investigate additional dynamical scenarios—such as buoyancy-driven oscillations and the stalling of binary SMBHs in rotating systems—to evaluate whether they can account for the observed photometric and kinematic features. These comparisons allow us to assess whether a recent recoil remains the most compelling explanation, or whether alternative processes could account for the observed off-centered nucleus.

\section{Alternative Dynamical Scenarios for Black Hole Displacement} \label{sec:altscen}

\subsection{Buoyancy-Induced Black Hole Offset} \label{sec:buoyancy}
%\SD{Shashank: If the theme of this section is ``mechanisms which can produce offsets but can't fully explain the data", then I think the buoyancy section should be here. But I'm also fine if you want to move it under the previous section.} \BT{I agree to leave it here}
Dynamical buoyancy is the opposite of dynamical friction, where the host system exerts a net positive (outward) torque on a massive perturber, leading to its outward motion \citep{Cole2012,Banik2021,Banik2022,Dattathri2025b}. This phenomenon is typically associated with cored density profiles. Since NGC~4486B has a central core (Figure~\ref{fig:3d_profile}), we investigate here whether dynamical buoyancy operates, and if so, what the associated offsets are. 

\begin{figure}
    \centering
    \includegraphics[width=\columnwidth]{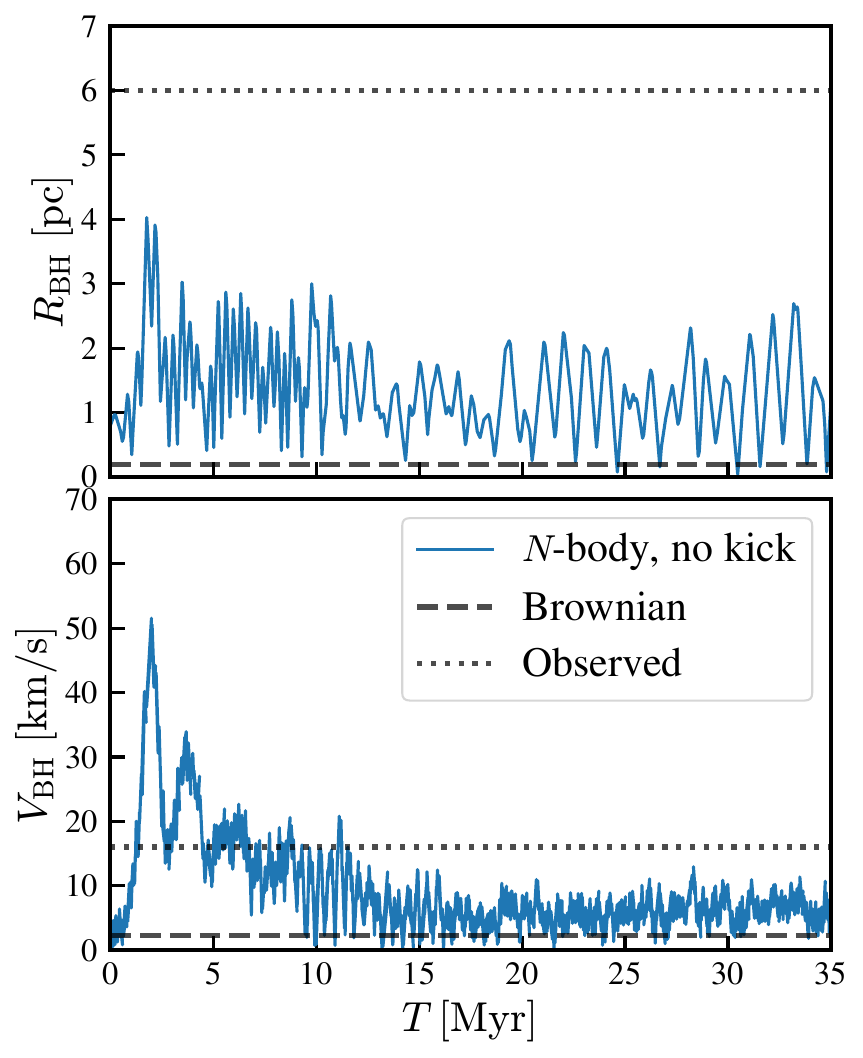}
    \caption{Trajectory of the black hole that is placed at the center of NGC~4486B \textit{without} any recoil kick. The black hole shows a clear outward motion away from the galactic center that is well resolved above the Brownian amplitude. However, except for very early times, the associated offsets are well below the observed values. }
    \label{fig:buoyancy}
\end{figure}

We initialize an $N$-body simulation with the same setup and parameters as Section~\ref{ssec:nbody_kick}, but \textit{without any recoil kick}, i.e., the SMBH is initially at rest at the galaxy center. We then let the system evolve and track the SMBH's motion around the galaxy center. Figure~\ref{fig:buoyancy} shows the resulting trajectory of the SMBH's position (top) and velocity (bottom). Similar to Figure~\ref{fig:trajectories}, the black dashed line shows the expected Brownian motion amplitude, and the black dotted line shows the observed offsets. The SMBH shows a clear initial outward motion even without a recoil kick. This motion is well resolved above the Brownian amplitude, indicating that it is not a numerical artifact. The offset reaches a maximum amplitude of $\sim 50\kms$  and $4$ pc away from the center at $T \sim 2$ Myr. However, similar to the recoil simulations (Figure~\ref{fig:trajectories}), the black hole quickly sinks back to the center. After $T \gtrsim 15$ Myr, the SMBH is orbiting around the galaxy center at a speed of $\sim 5$\kms and a distance $1-2$ pc, which is significantly lower than the observed offsets. 

The origin of the buoyancy in our system is unclear. \citet{Dattathri2025b} suggested that in spherical isotropic systems, dynamical buoyancy is due to an unstable dipole mode in the system. This instability may arise if the distribution function has a region where ${\rm d}f/{\rm d}E>0$ \citep{Dattathri2025a}. However, NGC~4486B is not isotropic and has a net rotation (Figure~\ref{fig:phot}); hence, it is unclear whether the same instability would occur here. A recent study by \citet{Boily2025} analyzed the motion of BHs in anisotropic galaxy cores and found that if the system has an excess of streaming motion over its velocity dispersion, a central black hole can gain angular momentum and move outward. Both of the above processes could work in tandem to produce an offset SMBH. However, regardless of the mechanism, it is clear from Figure~\ref{fig:buoyancy} that if the SMBH in NGC~4486B was initially at rest in the center, it could attain a significant velocity offset due to dynamical buoyancy for a short period of time $\sim 6-7$Myr, but the position offset is always well below the observed value of 6 pc. Note that the observed 6 pc separation is a projected distance; the true 3D offset can only be larger, which further strengthens this argument.

\subsection{Pre-Merger Binary Black Hole in Prograde Orbit}
\label{sec:premerger_binary}

\begin{figure*}
    \centering
    \includegraphics[width=\linewidth]{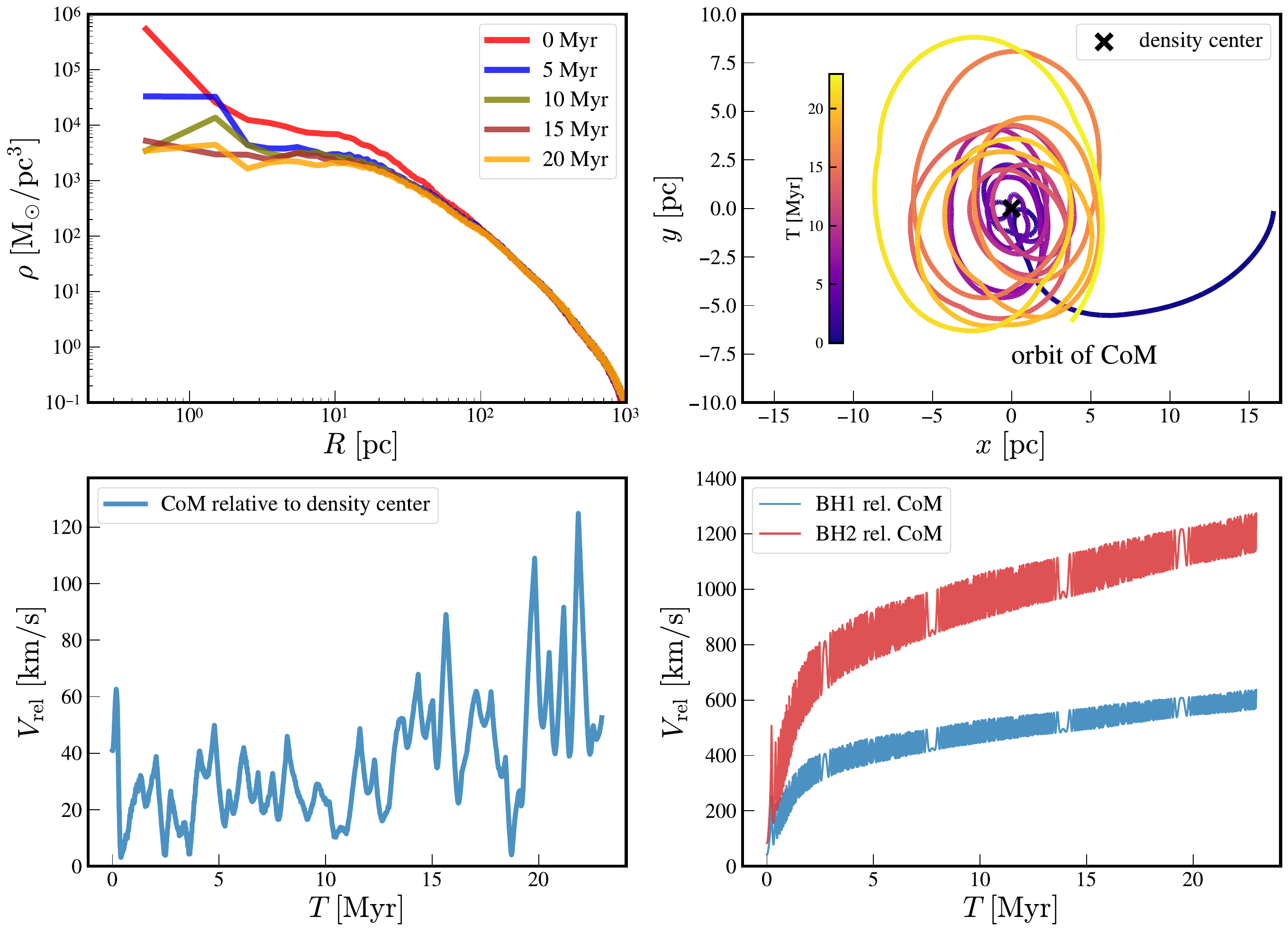}
    \caption{Time evolution of the galaxy and the SMBH binary parameters.
Top left: Volume-density profiles of the NGC 4486B model at different times during the SMBH binary evolution. Top right: Orbit of the SMBH binary in the reference frame of the stellar density center. The time-colored line shows the trajectory of the binary CoM after subtracting the density center motion; the density center itself is shown at the origin with a red dot. Bottom left: Velocity of the SMBH binary CoM relative to the density center, illustrating the residual motion of the binary within the host galaxy. Bottom right: Velocities of the individual SMBHs relative to the binary CoM, showing that the lighter secondary SMBH moves faster than the primary and that both velocities increase as the binary hardens.}
    \label{fig:binary}
\end{figure*}

Although NGC 4486B is overall a dispersion-dominated galaxy, it exhibits a considerable central rotation ($|v|/\sigma \simeq$ 0.40-0.65). Studies have shown that even this level of rotational support can profoundly influence the dynamics of SMBH binaries in rotating hosts \citep{holley+15,Rasskazov+16,Mirza+17,Khan2020}. In particular, simulations demonstrate that when a binary is on a prograde orbit aligned with the stellar rotation, its center-of-mass (CoM) can settle into a stable orbit around the galactic center with a characteristic radius comparable to the influence radius of the two BHs. Once established, such a system can survive for extended periods, maintaining a corotating orbit while the SMBHs in binary continue to shrink their relative separation, ultimately coalescing following GW emission.
%\mv{need to specify how long this is after the galaxy-galaxy merger with reference if we are to explain why the merger didn't happen in the outskirts of the cluster}, maintaining a corotating orbit rather than sinking immediately to the galaxy's center. 
%\mv{The abstract of \citet{holley+15} says that the binary on the prograde orbit merges 100Myrfaster than a binary in a non-rotating galaxy. How long does it take in a non-rotating galaxy? I}

This scenario is especially relevant for NGC 4486B. Photometric and kinematic studies indicate that it is an overall old, dynamically relaxed galaxy: its stellar populations are predominantly ancient, its large-scale light distribution is smooth, and no evidence exists for recent large-scale tidal disturbances \citep[e.g.,][]{Ferrarese.2006,Guerou.2015, Janz.2016}. Thus, any major galaxy–galaxy merger that contributed to its present structure likely occurred several gigayears ago \citep{Khan2018,Kelly2025}, consistent with its environment deep within the Virgo cluster. Nevertheless, if that galaxy–galaxy merger brought in a secondary SMBH on a prograde trajectory, the resulting binary could have settled into a long-lived corotating orbit before the SMBHs achieved a coalescence. This outcome naturally extends the effective lifetime of the binary off-center phase and is supported by the $N$-body simulations of rotating systems cited above.

To explore this possibility, we constructed a direct $N$-body model of NGC 4486B with one million particles, including the central SMBH with its mass from the best-fit Schwarzschild model (identical to that used in Section~\ref{ssec:nbody_kick}). A secondary SMBH, with half the primary’s mass, was introduced on a prograde orbit at a separation of 50 pc from the center with an initial velocity equal to 50\% of the circular velocity. The system was evolved with a GPU-accelerated $\phi$GPU code \citep{Berczik2013}. A bound binary formed after $\sim 5$ Myr, initiating a three-body scattering phase that flattened the inner density cusp into a shallow core (top panels of Figure~\ref{fig:binary}). During this phase, the binary settled into a stable orbit around the density center with a major axis of $\sim 10–15$ pc. Shortly after binary formation, the CoM velocity—measured relative to the density center—was modest $15–20$ km s$^{-1}$, but it subsequently exhibited mild fluctuations driven by continuing three-body interactions with surrounding stars (bottom-left panel of Figure~\ref{fig:binary}). In contrast, the velocities of the individual SMBHs—measured relative to the binary CoM—remained high and increased gradually as the binary hardened
 (bottom-right panel of Figure~\ref{fig:binary}),
These high internal velocities indicate that coalescence could still occur subsequently, with the merged SMBH inheriting the orbital parameters of the binary CoM. However, such an off-center binary configuration alone cannot reproduce the observed double nucleus morphology; forming an eccentric nuclear disk via a GW recoil kick remains the more plausible explanation. 

Once again, our initial $N$-body model reflects the current structure of NGC~4486B rather than that of its progenitor, which likely possessed a steeper central density cusp prior to binary scouring. Even so, the simulation demonstrates that in a host galaxy with the modest but non-negligible large-scale rotation observed, a forming SMBH binary can become trapped in a corotation resonance for a finite period. Because of the high computational cost of direct $N$-body integrations, we have evolved this model only to $\sim 30$ Myr. Nevertheless, our results indicate that while NGC 4486B is globally an old and relaxed system, its central SMBH binary could plausibly have remained in a long-lived resonant, corotating configuration following a galaxy–galaxy merger in the cluster outskirts (the Virgo cluster crossing time is $\sim10^9$ yr), and may have undergone coalescence only relatively recently. We estimated the supermassive black hole (SMBH) binary coalescence time using a semi-analytical prescription that accounts for orbital energy losses driven by both stellar dynamical interactions and GW emission \citep{khan+12a,sesana+15}. The stellar hardening rate was directly measured from our simulations and incorporated into the model to capture the contribution from three-body encounters. In addition, we included the energy and angular momentum losses due to GW radiation following the formalism of \citet{peters+63}, who derived the GW energy loss for two point masses in orbit about one another. Based on our estimates, the SMBH binary would require $\approx 500~\mathrm{Myr}$ to coalesce after its formation and this pahase will be preceeded by dynamical-friction–driven inspiral that itself may take several hundred Myr if the secondary SMBH arrives at a separation of $\sim 1~\mathrm{kpc}$ during the preceding galaxy–galaxy merger.

In a forthcoming study, we will explore this scenario using a larger suite of long-duration direct $N$-body simulations—initiated from a more strongly cusped pre-scouring density profile—and follow the evolution through GW–driven coalescence, with the goal of determining whether the merged SMBH can retain a sustained orbit around the galaxy’s density center.

\section{Summary and Conclusion} \label{sec:conclusion}

Although the double nucleus in NGC~4486B was first discovered nearly 30 years ago \citep{Lauer.1996} there have been few studies of its origin and dynamical structure. In analogy with the double nucleus in M31, \citet{Lauer.1996} proposed that it was an apsidally aligned eccentric nuclear disk surrounding an SMBH. Two striking differences between M31 and NGC~4486B however are: (a) in M31 the SMBH (located at P3) is at the galactic center while the SMBH in NGC~4486B is 6 pc away from the center (defined by the large scale isophotes) and possibly has a velocity offset of $\sim 16\kms$ relative to stars on the opposite side of the galaxy; (b) M31 has a central stellar cusp and a relatively young star cluster coincident with the SMBH while  NGC~4486B has a flat core $\gamma=0.14$ with a core radius of $\sim 20$ pc. In Paper 1 \citep{Behzad.2025} we used JWST-NIRSpec IFU spectroscopy and archival HST photometry combined with orbit-superposition Schwarzschild dynamical modeling to confirm the presence of an SMBH of $\mbh = 3.6\pm10^8\msun$. Despite the fact that the Schwarzschild model is axisymmetric and consequently does not reproduce the asymmetric photometry and kinematics, the distribution function provided by the model allows us to examine the origin of the double nucleus and the offset SMBH in NGC~4486B for the first time. 

In this work, we focus on the possibility that the double nucleus in NGC~4486B is due to an END similar to that in M31 \citep{Tremaine.1995}. We compare the orbital properties of the Schwarzschild model with predicted properties of ENDs that are produced when an SMBH surrounded by a circular nuclear disk experiences a GW recoil kick. These simulations show that a GW recoil following SMBH coalescence can transform an initially circular stellar disk into an eccentric, apsidally aligned structure with a significant retrograde population. Previous studies of surface brightness and kinematic distributions of simulated ENDs match many properties of the observed double nucleus in NGC~4486B: the fainter peak and the velocity dispersion peak coincide and are the location of the SMBH, the asymmetric line-of-sight velocity field, are all consistent with predictions for nearly edge-on ENDs. This interpretation is further reinforced by the finding that our best-fitting Schwarzschild models require a significant fraction of retrograde orbits in the outer nuclear region. Retrograde orbits are not only a natural dynamical consequence of a recoiling SMBH but also a key ingredient in the long-term stability of ENDs. The presence of the central core in this galaxy is additional evidence that it has experienced a binary-black hole mediated ``cusp scouring'' prior to the binary black hole merger that likely gave the resulting SMBH the kick which produced the END.

From the observed properties of the double nucleus  (semi-major axis $\sim 6$~pc) and black hole mass we estimate that the SMBH experienced a recoil kick of $\sim 340\kms$. We adapt the formalism of \citep{Akiba.2025} to estimate the recoil kick magnitude as a function of the pre-merger black hole mass ratio, $q=m_2/M_1$.  Marginalizing over the spin distribution of the two BHs (which is assumed to be isotropic in direction and uniform in scaled magnitude between [0,1]), we estimate that for a recoil kick of $340\kms$, the pre-merger mass ratio was: $q \sim 0.24\pm0.09$ implying that the mass of the secondary was $m_2 \sim 7\pm2.3\times10^7\msun$. 

To explore this scenario further, we constructed $N$-body realizations of our best-fit Schwarzschild model from Paper 1 and carried out numerous experiments, which show that a recoiling SMBH, kicked with a velocity of $0.2-0.75 v_{\rm esc} \sim 154-576\kms$, oscillates and sinks back to the galactic center within $\sim$10-80~Myr. For our estimated recoil kick of $340\kms$, the timescale to sink to the center is expected to be less than 30~Myr, implying that if the position and velocity offsets of the SMBH are to be attributed to a GW recoil kick, the binary black hole merger in NGC~4486B must have occurred quite recently. The simulations rule out ``core stalling'' and ``dynamical buoyancy'' scenarios in which the failure of dynamical friction in a constant density core would allow the SMBH to oscillate for extended periods \citep{Dattathri2025b}. This discrepancy is likely due to the overmassive nature of the SMBH in NGC~4486B ($\mbh/M_*\sim 4-10$\%), while most numerical simulations of this problem assume $\mbh/M_*\sim 0.1-0.2$\%, based on expectations of SMBH scaling relations. 

%We also tested alternative explanations for the observed offsets. Simulations of buoyancy-driven oscillations and Brownian motion show that while these processes can induce displacements of a central SMBH, the amplitudes fall far below the observed values. Buoyancy produces somewhat larger excursions than Brownian motion, but still underpredicts the velocity and spatial offsets of NGC~4486B at late times. Thus, neither mechanism can account for the kinematic asymmetry, reinforcing recoil as the most plausible explanation. %Importantly, our models also demonstrate that the $\sim$16 km s$^{-1}$ LOS velocity enhancement at the dispersion peak should not be interpreted as the bulk velocity of the SMBH itself, but rather as the mean motion of stars bound to the SMBH in the eccentric disk. This subtlety emphasizes the need to disentangle stellar streaming signatures from true black hole motion.

%Our inference that the recoil event was very recently present presents a paradox: the proximity of NGC~4486B to the center of the Virgo Cluster and the fact that it is has a uniformly old stellar population, which appears dynamically relaxed. The lack of tidal features and the uniformly old stellar population indicate that the $\sim 1:4 -1:5$ major merger that brought in the secondary BH must have taken place several gigayears ago, while the two galaxies were in the outskirts of the Virgo cluster.  

Our inference that the recoil event occurred recently presents an apparent paradox. NGC 4486B lies near the dense center of the Virgo Cluster and exhibits a uniformly old, dynamically relaxed stellar population. The absence of tidal features and the uniformly old stars imply that the $\sim 1:4 -1:5$ major galaxy–galaxy merger that introduced the secondary black hole must have taken place several gigayears ago, likely when the progenitor galaxies were still in the cluster outskirts.

%NGC~4486B is an elliptical galaxy with significant rotational support $|v|/\sigma = 0.45-0.6$.   Our direct $N$-body simulations of SMBH binaries in a rotating host galaxy (constructed from our best-fit Schwarzschild model)  provides insight. We find that when the BH binary forms on a prograde orbit with angular momentum aligned with the galaxy’s rotation, the binary's center of mass can settle in a long-lived corotation resonance at scales comparable to its influence radius. Such binaries may persist for extended periods before coalescing, before eventually merging (roughly 500 Myr after a SMBH binary forms for this case) and imparting a recoil to the remnant SMBH. Our simulations indicate that the binary BH merger can occur off-center ($\sim$10–15 pc) with a center-of-mass velocity of $15–20\kms$. However, this binary-merger scenario alone cannot reproduce the distinct double-nucleus morphology. The formation of an END remains the more plausible explanation for the photometric structure.

NGC~4486B is an elliptical galaxy with significant rotational support ($|v|/\sigma = 0.45-0.65$). Our direct $N$-body simulations of SMBH binaries embedded in a rotating host galaxy—constructed using our best-fit Schwarzschild model—offer additional insight. When the binary forms on a prograde orbit aligned with the galaxy’s rotation, its center of mass can become trapped in a long-lived corotation resonance at scales comparable to the black hole’s sphere of influence. Such binaries may persist for extended periods before coalescing, before eventually merging (roughly 500 Myr after an SMBH binary forms for this case) and imparting a recoil to the remnant SMBH. Our simulations indicate that the SMBH merger can occur off-center ($\sim$10–15 pc) with a center-of-mass velocity of $15$–$20\,\mathrm{km \, s^{-1}}$. However, this binary-merger scenario alone cannot reproduce the distinct double-nucleus morphology. The formation of an END remains the more plausible explanation for the observed photometric structure, reinforcing our conclusion that the black hole merger occurred recently ($\sim$ 30 Myr), despite the galaxy’s otherwise ancient and dynamically relaxed stellar population.

The implications of this finding extend beyond a single galaxy. If compact ellipticals with double nuclei arising from ENDs like NGC~4486B retain signatures of relatively recent SMBH mergers, they may provide valuable laboratories for studying post-merger dynamics on resolvable scales. %Systems like this may provide a crucial connection between local nuclear dynamics and the unresolved population of SMBH binaries contributing to the nanohertz gravitational-wave background recently detected by pulsar timing arrays \citep{Nanograv.2023, EPTA.2023}. 
Non-equilibrium dynamical models similar to those constructed for M31 \citep{ Peiris.2003, Bacon.2001, Sambhus.2002, BrownMagorrian.2013} fitted to the excellent NIRSpec-IFU kinematic maps of NGC~4486B and $N$-body models of the resulting END, embedded inside the rotating elliptical host are needed to more precisely determine the time since the SMBH merger event and the expected survival time of the END. NGC~4486B serves not only as a rare nearby case study, but also as a cornerstone for linking high-resolution observations of galactic centers to GW astrophysics.

\begin{acknowledgements}
We are grateful to Tod Lauer for kindly providing the deconvolved F555W image from \citep{Lauer.1996}. We thank Eugene Vasiliev for
useful discussions. BT and MV acknowledge funding from Space Telescope Science Institute awards: JWST-GO-02567.002-A and HST-GO-16882.002-A.  M.A.T.\ and S.T.\ acknowledge the support of the Canadian Space Agency (CSA) [22JWGO1-07]. This work was supported by Tamkeen under the NYU Abu Dhabi Research Institute grant CASS, and utilized the high-performance computing resources at New York University Abu Dhabi. CL acknowledges support from the National Natural Science Foundation of China (NSFC, Grant No. 12173025), National Key R\&D Program of China (2023YFA1607800, 2023YFA1607804), 111 project (No. B20019), and Key Laboratory for Particle Physics, Astrophysics and Cosmology, Ministry of Education. B.T. acknowledges support from the Mendel Science Experience (MSE) Postdoctoral Fellowship awarded by Villanova University. MV and BT acknowledge support from NASA through grant number HST-AR-18142 awarded by the Space Telescope Science Institute, which is operated by AURA, Inc., under NASA contract NAS 5-26555.

This work is based on observations made with the JWST and obtained from the Mikulski Archive for Space Telescopes (MAST). The data are available at \dataset[doi:10.17909/4rr8-zk52]{https://doi.org/10.17909/4rr8-zk52}.

\facilities{JWST(NIRSpec), HST(ACS/WFC), HST(WFPC2)}

\end{acknowledgements}

\bibliography{1.bibtex}{}
\bibliographystyle{aasjournal}

\appendix
\setcounter{figure}{0} % Reset figure counter
\renewcommand{\thefigure}{A\arabic{figure}} % Change numbering format

 \begin{figure*}
	\centering	%
	\includegraphics[width=0.95\columnwidth]{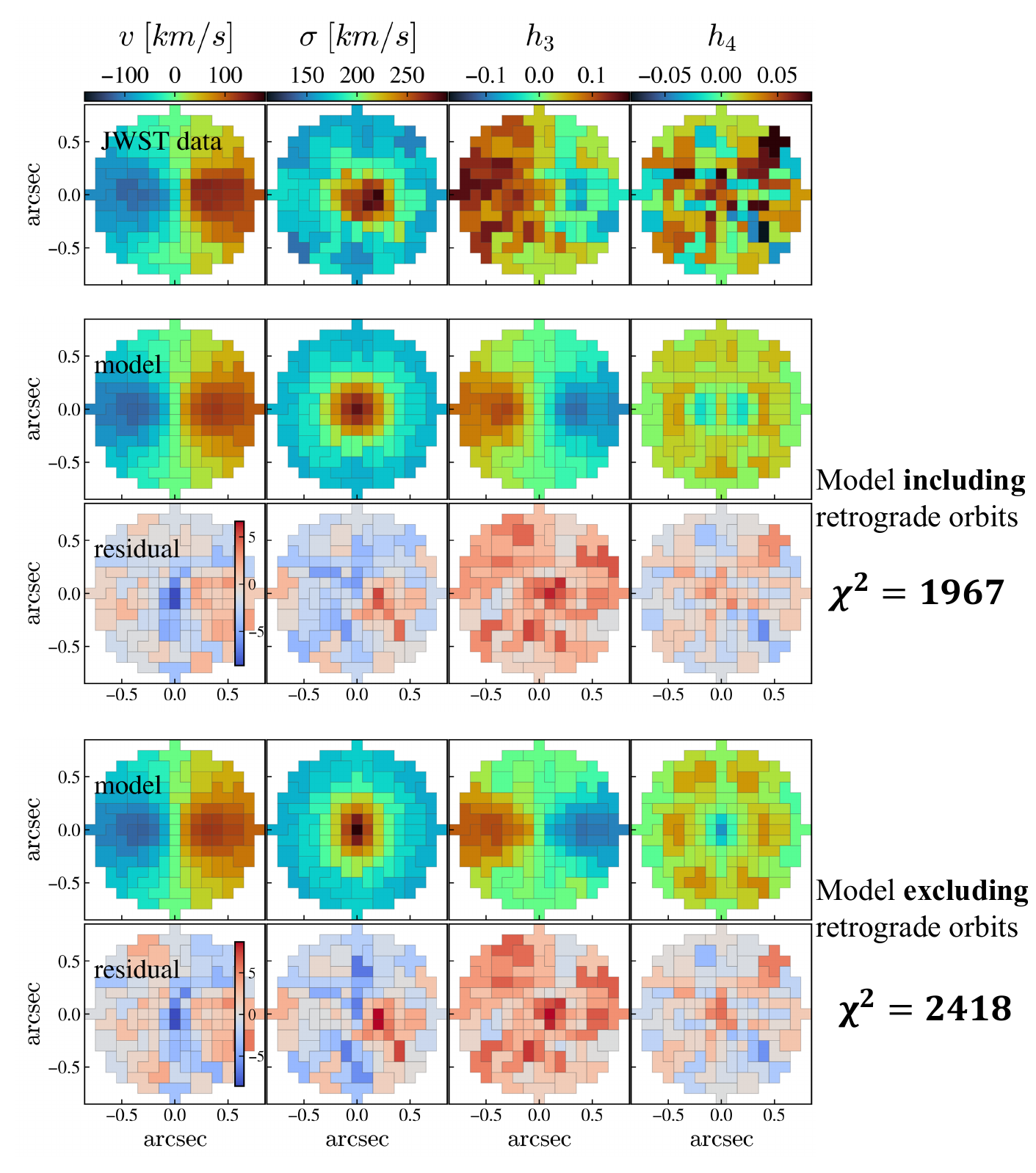}
	\caption{Best-fit Schwarzschild models with (top) and without (bottom) retrograde orbits, each annotated with the corresponding $\chi^2$ values.}%
	\label{fig:Schw_Ap}%
\end{figure*}

\clearpage

\end{document}